\documentclass[11pt]{article}
\pdfoutput=1
\usepackage{jheparxiv}

\usepackage[dvipsnames]{xcolor}

\usepackage{tikz-feynman}
\usepackage{amsmath}
\usepackage{amsfonts}
\usepackage{bm}
\usepackage{braket}
\usepackage[normalem]{ulem}
\usepackage{comment}
\usepackage{mathrsfs}  
\usepackage{subcaption}
\usepackage{orcidlink}

\usepackage{hyperref}
\hypersetup{colorlinks= true, linkcolor=blue, citecolor=red}

\newcommand{\be}{\begin{equation}}
\newcommand{\ee}{\end{equation}}
\newcommand{\ud}{\mathrm{d}}

\newcommand{\LCm}{{\scriptscriptstyle -}} %LC supersripts
\newcommand{\LCp}{{\scriptscriptstyle +}}
\newcommand{\LCpm}{{\scriptscriptstyle \pm}}

\newcommand{\LCperp}{{\scriptscriptstyle \perp}}
\newcommand{\ii}{{\mathbf{i}}}
\newcommand{\jj}{{\mathbf{j}}}

\title{Tunnelling amplitudes and Hawking radiation from worldline QFT}
\author{Anton Ilderton \orcidlink{0000-0002-6520-7323}}
\emailAdd{anton.ilderton@ed.ac.uk}
\author{and Karthik Rajeev \orcidlink{0000-0003-3193-1900}}
\emailAdd{karthik.rajeev@ed.ac.uk}

\affiliation{Higgs Centre, School of Physics and Astronomy, University of Edinburgh, EH9 3FD, UK}

\date{}
\abstract{
We compare Hawking radiation in a collapse background with Schwinger pair creation in an electric field.
The comparison is driven by the presence of an analogue horizon in the Schwinger case, which causally divides spacetime for classical particles, but through which quantum fields can tunnel.
Amplitudes for tunnelling processes are 
encoded in the asymptotic behaviour of solutions to the appropriate background-coupled wave equation.
We construct these solutions, in both gravity and QED, using the worldline approach, where tunnelling and particle creation manifest as complex saddle points of a real-time path integral.
For the Schwinger effect, these saddles correspond to complex worldlines, while for Hawking radiation the corresponding worldlines are real, but appear complex when extended beyond a certain coordinate patch.
}
\begin{document}

\setcounter{tocdepth}{2}
\maketitle

\section{Introduction}
%%%%%
The application of on-shell methods to extract classical observables from amplitudes has emerged as an effective method for analysing gravitational physics, for recent reviews see~\cite{Travaglini:2022uwo,Kosower:2022yvp,Buonanno:2022pgc}. As well as providing direct information on the classical gravitational waves emitted in two-body scattering events~\cite{Kosower:2018adc,Cristofoli:2021vyo}, the amplitudes-based formalism also offers insights into e.g.~bound dynamics~\cite{Kalin:2019rwq,Kalin:2019inp,Cho:2021arx,Adamo:2022ooq,Gonzo:2023goe,Adamo:2024oxy,Buonanno:2024byg}, tidal response~\cite{Aoude:2020ygw,Ivanov:2022qqt,Jakobsen:2023pvx,Saketh:2023bul}, and supertranslations~\cite{Georgoudis:2023eke,Bini:2024rsy,Elkhidir:2024izo}, to name but a few. Scattering amplitudes and other QFT tools have naturally also found application to gravitational problems in the semiclassical regime~\cite{Goldberger:2020geb,Kim:2020dif,Gaddam:2021zka,Gaddam:2020mwe,Ferreira:2020whz,Melville:2023kgd}, including semiclassical evaporation of black holes through Hawking radiation~\cite{Aoude:2024sve}.

One of the most prominent approaches to Hawking radiation describes it in terms of the quantum mechanical tunnelling of states through the horizon~\cite{Hartle:1976tp,Damour:1976jd}. The related `complex-path analysis' approach~\cite{Srinivasan:1998ty} reformulates tunnelling in terms of a Hamilton–Jacobi action, offering a manifestly covariant perspective~\cite{Shankaranarayanan:2000gb} and a systematic extension to more general black hole spacetimes~\cite{Shankaranarayanan:2000qv,Padmanabhan:2004tz,Shankaranarayanan:2003ya,Angheben:2005rm,Kerner:2006vu}. The tunnelling approach has also been extended to account for leading effects of backreaction on the spacetime geometry due to created pairs~\cite{Parikh:1999mf}, and to the incorporation of a cosmological constant and higher-dimensional black holes~\cite{Hemming:2000as,Volovik:2008ww,Parikh:2002qh,Medved:2002zj,Wu:2006nj}.

Another extensively studied example of tunnelling is pair creation in an electric field, the (famously non-perturbative) `Sauter-Schwinger effect'~\cite{Sauter:1931zz,Schwinger:1951nm}.  In this context tunnelling does not strictly denote potential barrier penetration in the sense of non-relativistic quantum mechanics; while the pair creation rate is analogous to either over-the-barrier reflection or tunnelling through a barrier, depending on the gauge used~\cite{Brezin:1970xf,Srinivasan:1998ty}, the use of the term `tunnelling' is ultimately justified by the presence of semi-classical, exponentially suppressed factors, similar to those found in quantum mechanics. There are many extensions of the Schwinger effect beyond the constant-field case, see~\cite{Fedotov:2022ely} for a review and further references.

The Schwinger effect is conceptually simpler than Hawking radiation, and may seem only qualitatively related, primarily because there is no event horizon to deal with -- except that both classical and quantum dynamics in a background electric field can be phrased in terms of what happens at an analogue horizon, within electromagnetism~\cite{Srinivasan:1998ty,
Srinivasan:1999ux,Ilderton:2023ifn}, suggesting a closer connection than may initially be expected.

In this paper we will explore how amplitudes-based approaches to both the Schwinger effect and Hawking radiation are connected to the tunnelling picture. We will make this connection through the worldline formulation of field theory~\cite{AFFLECK1982509,Bern:1990cu,Bern:1991aq,Strassler:1992zr,Edwards:2019eby}, in which Feynman diagrams are traded for first quantised (particle) path integrals, for a review see~\cite{Edwards:2019eby}. This approach is particularly useful for generating all-multiplicity master formulae for amplitudes~\cite{Daikouji:1995dz,Martin:2003gb,Schubert:2001he,Dunne:2002qf,Ahmadiniaz:2020wlm,Ahmadiniaz:2021gsd}, and has been applied to the casimir effect~\cite{Gies:2003cv,Gies:2006bt}, amplitudes in backgrounds~\cite{Shaisultanov:1995tm,Reuter:1996zm,Schubert:2000yt,Bastianelli:2002fv,Edwards:2021vhg,Schubert:2023gsl,Copinger:2023ctz,Copinger:2024twl} and quantum gravity~\cite{Bonezzi:2018box,Bastianelli:2019xhi,Bastianelli:2013tsa}. Worldline QFT has recently been applied to problems in classical gravity, such as light-bending~\cite{Bastianelli:2021nbs} and, in particular, binary dynamics \cite{Mogull:2020sak,Jakobsen:2021smu,Jakobsen:2021lvp,Jakobsen:2021zvh,Jakobsen:2022fcj,Haddad:2024ebn,Driesse:2024feo}.

Connections between the worldline formalism and the tunnelling picture have been explored from various angles, but investigations beginning from an explicitly real-time (rather than Euclidean) path integral remain limited, see e.g.~\cite{Ilderton:2014mla,Feldbrugge:2019sew,Rajeev:2021zae}. To the best of our knowledge, the \textit{Lorentzian} worldline approach has not yet been extended to the case of Hawking radiation. We will bridge this gap here.

This paper is organised as follows. In Sec.~\ref{sec:tunnel_EM} we discuss the classical and quantum physics of a (scalar) particle in a constant electric field, with the presentation adapted to explicitly emphasise the electromagnetic horizon and similarities to gravitational particle production. We will see that the worldline formalism provides a robust framework for constructing solutions of background-coupled wave equations, which are the basic building blocks of scattering amplitudes in curved spacetime and background fields~\cite{Furry:1951bef,DeWitt:1967ub,tHooft:1975uxh,Abbott:1981ke}. This study will also motivate choices made in our subsequent discussion of Hawking radiation.

In Section~\ref{sec:modes_amplitudes} we connect the Bogoliubov approach to pair creation to the amplitudes-based approach. Equipped with this and with useful insights from the Schwinger effect, we turn to Hawking radiation in Sec.~\ref{sec:hawking}. We begin by reviewing classical trajectories in the Vaidya metric representing radial collapse of an infinitesimally thin spherical null shell.  We then give the Lorentizian path integral formulation for constructing mode functions in the Vaidya spacetime, and use this to recover  standard (Bogoliubov type) results for Hawking radiation~\cite{Hawking:1975vcx}. We finally construct tunneling wavefunctions from which we read off the particle creation amplitude for Hawking radiation directly. We conclude in Sec.~\ref{sec:conclusion}. The appendices contain additional worldline amplitude calculations.

\section{Tunnelling in electromagnetic fields}\label{sec:tunnel_EM}
%%%%%%%%
In this section we explore tunnelling in electric fields. We will work with the simplest case of a constant field, which is sufficient to reveal several similarities with Hawking radiation. For related discussions and the non-constant case see~\cite{Srinivasan:1999ux,Tomaras:2000ag,Tomaras:2001vs,Ilderton:2023ifn}.

%%%%%%%%%%%%%%%%%%%%%%%%%%%%%%%%%%%%%%%
\subsection{Physics at effective horizons}\label{subsec:classical_EM}
%%%%%%%%%%%%%%%%%%%%%%%%%%%%%%%%%%%%%%%
The physics of interest, in particular the analogy with gravitational horizons, is made most explicit by working in lightfront coordinates,
\be
    \ud s^2=2\ud x^{\LCp}\ud x^{\LCm}- \ud x^\LCperp \ud x^\LCperp \;.
\ee
We define null vectors $\{n_\mu, {\bar n}_\mu\}$ by $n\cdot x=x^\LCm$, ${\bar n}\cdot x = x^\LCp$, ${\bar n}\cdot n = 1$.
Our electric field then has a field strength tensor with $F_{\LCm\LCp}= E$ and all other components vanishing. The Lorentz force equations of motion for a particle in this field reduce to
\be\label{eq:classical-eom}
    m \ddot{x}^\LCpm = \pm eE {\dot{x}}^\LCpm \;, \quad {\ddot x}^\LCperp = 0 \;.
\ee
Writing $\pi_\mu = m \dot{x}_\mu$ for the kinematic momenta, the first integral of (\ref{eq:classical-eom}) is
\be\label{eq:classical-momenta}
    \pi_\LCp = p_\LCp - e E x^\LCm \;,\quad \pi_\LCm = p_\LCm + e E x^\LCp \;, \qquad \pi_\LCperp = p_\LCperp \;,
\ee
where for simplicity we have chosen initial conditions $\pi_\mu(0) = p_\mu$ and $x^\mu(0)=0$. Example solutions are shown as part of Fig.~\ref{fig:EM-horizon}, below.

The statement that massive particles have speed less than $c$ translates to $\pi_\LCpm >0$ in lightfront coordinates. It is then apparent from (\ref{eq:classical-momenta}) that particles with $eE>0$ are accelerated to $c$ in finite `lightfront time' $x^\LCm$ (but infinite `instant form' time $x^0$); they cannot cross the surface
\be
    x^\LCm = x^\LCm_h := \frac{p_\LCp}{eE} \;.
\ee%
This null surface can be thought of as an electromagnetic `horizon'~\cite{Srinivasan:1999ux,Ilderton:2023ifn}, as the particle is causally disconnected from the region $x^\LCm>x^\LCm_h$. The analogy with gravitational horizons is not exact, since particles of the opposite charge can pass the horizon without issue -- however, these particles have an equivalent horizon at $x^\LCp = -p_\LCm/(eE)$. 

Turning to the quantum theory, we solve the Klein-Gordon equation $(D^2+m^2)\varphi=0$ to construct particle wavefunctions $\varphi$, where $D_\mu=\partial_\mu + ieA_\mu$. A convenient gauge potential for the constant electric field which manifests the connection to the classical physics above is
\begin{align}\label{eq:potential}
    eA_{\mu}= {\bar n}_\mu eE x^\LCm \;.
\end{align}
A natural ansatz for \emph{incoming} particle wavefunctions is then
\begin{align}
    \varphi_{p}(x)=e^{-ip_\LCp x^\LCp -ip_{\LCperp}x^{\LCperp}}f_{p}(x^{\LCm}) \;,
\end{align}
on which the KG equation reduces to the first-order form
\begin{align}\label{eq:KG-simplified}
   2 i p_{\LCp}\Big(1-x^{\LCm}/x^\LCm_h\Big) f_p'(x^{\LCm}) = \big(p_{\LCperp}^2 +m^2+i eE\big) f_p(x^\LCm) \;.
\end{align}
This would be trivial to solve except that the solution is ambiguous exactly at the location of the classical horizon $x^\LCm = x^\LCm_h$. To be precise, we write down the naive solution of (\ref{eq:KG-simplified}) in the region $x^\LCm<x^\LCm_h$, that is
\begin{align}\label{eq:incoming-basic}
    \varphi_p(x) = e^{-ip_\LCperp x^\LCperp -ip_\LCp x^\LCp}\, e^{-\left(\frac{1}{2}-i \frac{p^2_{\LCperp}+m^2}{2eE}\right)\log\left(1-\frac{x^{\LCm}}{x^\LCm_h}\right)}    \;.
\end{align}
The questions to answer are then `how do we continue this wavefunction past the horizon?', and `what is the associated physics?' It is here that the worldline approach is useful. 

%%%%%%%%%%%%%%%%%%%%%%%%%%%%%%%%%%%%%%%
\subsection{Worldline representation of particle wavefunctions}\label{subsec:WL_EM}
%%%%%%%%%%%%%%%%%%%%%%%%%%%%%%%%%%%%%%%
To construct particle wavefunctions in the worldline approach, we first recall the worldline representation of the scalar field propagator in a background $A_\mu$, that is
\be\label{prop-worldline-def}
    G(x,y) = \int\limits_0^\infty\!\ud T \, \int\limits_{z(0)=y}^{z(T)=x}\!\mathcal{D}z\, e^{i S}  \;,
    \qquad
    S =  -\int_0^T\!\ud\tau\,\bigg(\frac{1}{4}{\dot z}^2 + m^2 + e {\dot z} \cdot A(z)\bigg) \;,
\ee
in which $T$ is proper time. The path integral is taken over all worldlines $z^\mu(\tau)$ with Dirichlet boundary conditions at $\tau=0$ and $\tau=T$, and represents the quantum mechanical transition element between states $\ket{x}$ and $\ket{y}$ with Hamiltonian $H= D^2+m^2$, i.e.
\be
 G(x,y) \equiv \int\limits_0^\infty\!\ud T \, \bra{x} e^{-i (H-i\epsilon) T}\ket{y} \;.
\ee
We could use some other boundary condition at, say, $T=0$, or sew the propagator to some chosen state, replacing $\ket{y} \to \ket{\psi}$. The resulting path integral simply describes how information in $\ket{\psi}$ is propagated to $x^\mu$.
Now, observe that by computing the same path integral, but extending the proper-time integral to the entire real line, we no longer propagate but rather \emph{project} onto solutions of the Klein-Gordon equation since, formally,
\be\label{sol-worldline-def}
   \int\limits_{-\infty}^\infty\!\ud T \bra{x} e^{-i H T}\ket{\psi} =  \bra{x} 2\pi\delta(H)\ket{\psi} \;.
\ee
Let us apply this to our system of a scalar particle in an electric field. To describe the incoming wavefunction (\ref{eq:incoming-basic}), it is natural to impose initial conditions in momentum space, which translates to the choice $\braket{y|\psi}\propto e^{-ip\cdot y}$. With this, \eqref{sol-worldline-def} motivates us to consider the path integral
\be\label{eq:integral-for-wavefunction0}
\begin{split}
      \varphi_p(x) &:=
    \int\!\ud^4 y\, e^{-ip\cdot y} \int_{-\infty}^{\infty}\!\ud T\!\int^{z(T)=x}_{z(0)=y}\mathcal{D}z\,
    e^{iS} \;,
\end{split}  
\ee
with $p_\mu$ on-shell; attaching our chosen state amounts to taking a Fourier transform with respect to the initial worldline point $y^\mu$. We may convert (\ref{eq:integral-for-wavefunction0}) to the equivalent form 
\be\label{eq:integral-for-wavefunction}
\begin{split}
      \varphi_p(x) &=\int_{-\infty}^{\infty}\!\ud T\!\int^{z(T)=x}\mathcal{D}z\,
    e^{iS_{\text{WL}}}\;.
\end{split}  
\ee
in which we have free boundary conditions at $\tau=0$, but the worldline action acquires a boundary term, i.e.
\begin{align}\label{eq:worldline-action}
    {S}_{\rm WL}=-p\cdot z(0)-\int_{0}^{T}\!\ud \tau\, \bigg(\frac{\dot{z}^2}{4}+m^2 + eA\cdot \dot{z}\bigg) \;.
\end{align}
Practically, the simplest way to compute the path integral in (\ref{eq:integral-for-wavefunction})  is to use Dirichlet boundary conditions on both ends of the line, and then Fourier transform to momentum space. Since the integral is Gaussian, and well-known in closed form (for a recent treatment see~\cite{Ahmad:2016vvw}) we simply quote results. The exact form of the kernel $\braket{x|e^{-iHT}|y}$ in the gauge (\ref{eq:potential}) is~\cite{Srinivasan:1999ux}
\begin{align}
    \braket{x|e^{-iHT}|y}&=\frac{e^{-im^2T}}{16i\pi^2T}\frac{eE}{\sinh(eET)}e^{ \frac{i(x^{\LCperp}-y^{\LCperp})^2}{4T}}e^{-\frac{i eE (x^\LCp-y^\LCp) (x^\LCm-y^\LCm e^{-2
   eE T})}{\left(1-e^{-2 eE T}\right)}} \;.
\end{align}
The Fourier transform of the above is again Gaussian, and we obtain
\begin{align}\label{integral-to-do}
    \varphi_{p}(x)=\int_{-\infty}^{\infty}\!\ud T\, e^{eET} e^{iS_{\rm cl}} \;,
\end{align}
in which 
\begin{align}\label{eq:SCL}
    S_{\rm cl}=\frac{p_\LCm p_\LCp \left(e^{2 eE T}-1\right) (1-x^\LCm /x^\LCm_h)}{eE}-p\cdot x-(p_{\perp}^2+m^2)T \;.
\end{align}
Note that $S_{\rm cl}$ is the classical action, i.e. $S_{\rm WL}$ in~(\ref{eq:worldline-action}) evaluated on the solution of the classical worldline equations of motion obeying the boundary conditions
\be
    z^\mu(T)=x^\mu \;, 
    \qquad
    \tfrac12 \dot{z}^\mu(0)+eA^\mu(z(0))=p^\mu \;.
\ee
The solution is, explicitly, $z_{\rm cl}^{\LCperp}(\tau)=x^{\LCperp}-2p_{\LCperp}(T-\tau)$ and 
\begin{align}\label{eq:z_cl_tau}
    z_{\rm cl}^{\LCm}(\tau)&=x_h^\LCm +\left(x^{\LCm}-x^\LCm_h\right)e^{2eE(T-\tau)} \;, \qquad
     z_{\rm cl}^{\LCp}(\tau)=x^{\LCp}+\left(\frac{e^{2eE\tau}-e^{2eET}}{eE}\right)p_{\LCm} \,.
\end{align}
We note the appearance of the same factor $1-x^\LCm/x^\LCm_h$ in  (\ref{eq:SCL}) as in the wavefunctions of Sec.~\ref{subsec:classical_EM}. The action and integrand of (\ref{integral-to-do}) are, though, well-behaved everywhere, so resolution of what happens at the horizon must lie in evaluation of the $T$-integral, to which we now turn.

\subsection{Saddle point analysis}
%%%%%%%%%%%%%%%%%%%
%
We now analyse the $T$-integral (\ref{integral-to-do}), identifying the saddle points and steepest descent contours in the complex $T$--plane. We observe first that a change of variables $Y=e^{eE T}$ absorbs the leading exponential in (\ref{integral-to-do}) into the measure,  hence we may focus on the behaviour of the classical action $S_\text{cl}$. We continue to use $T$ for simplicity.

\begin{figure}[t!!]
    \centering
%%%%%%%%%%%%%%%   
    \begin{subfigure}[t]{0.45\textwidth}
    \includegraphics[width=\textwidth]{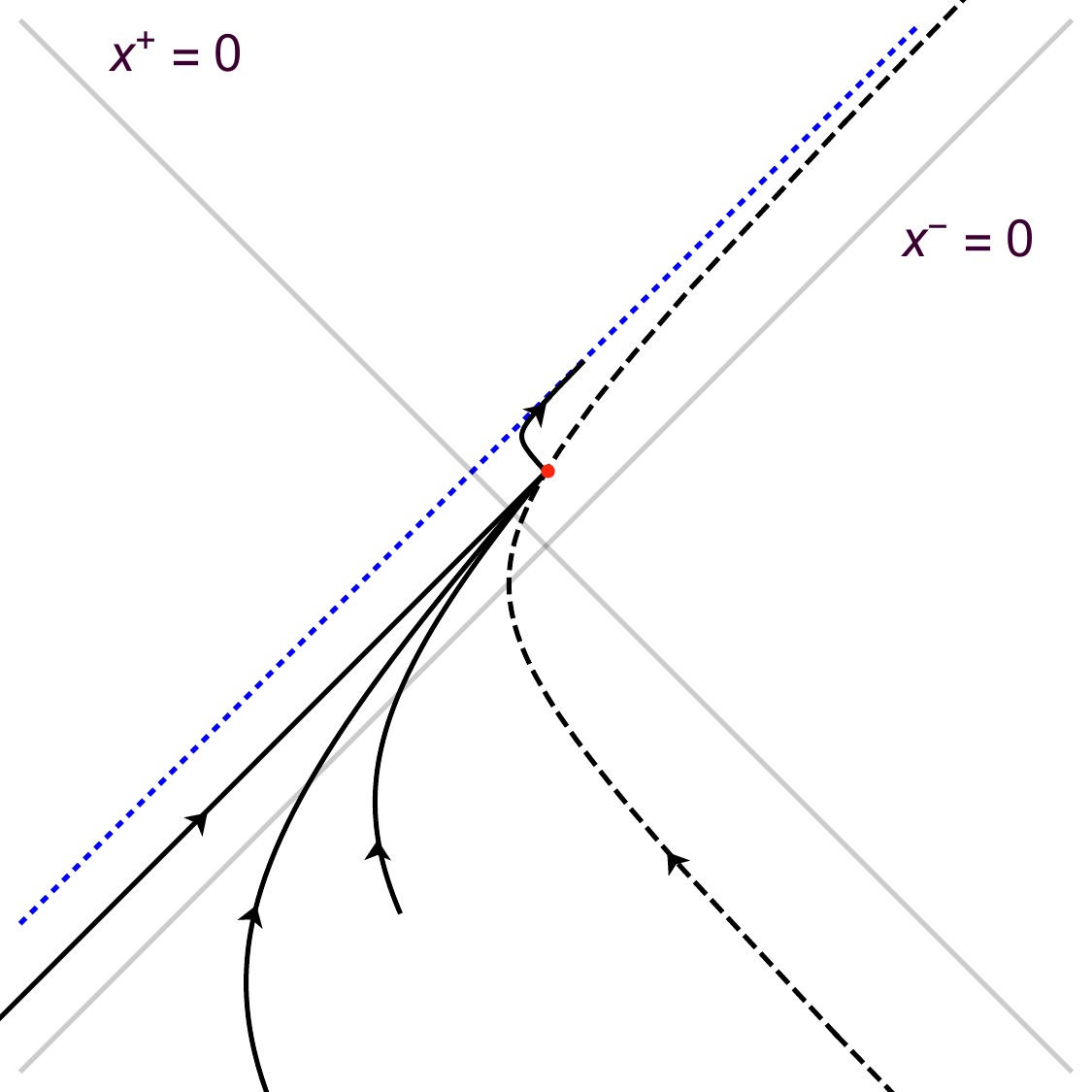}
\caption{$x^\LCm < x^\LCm_h$}
    \label{fig:WL_real-saddle}
    \end{subfigure}
%%%%%%%%%%%%%%
     \hfill
 %%%%%%%%%%%%%    
    \begin{subfigure}[t]{0.45\textwidth}
    \includegraphics[width=\textwidth]{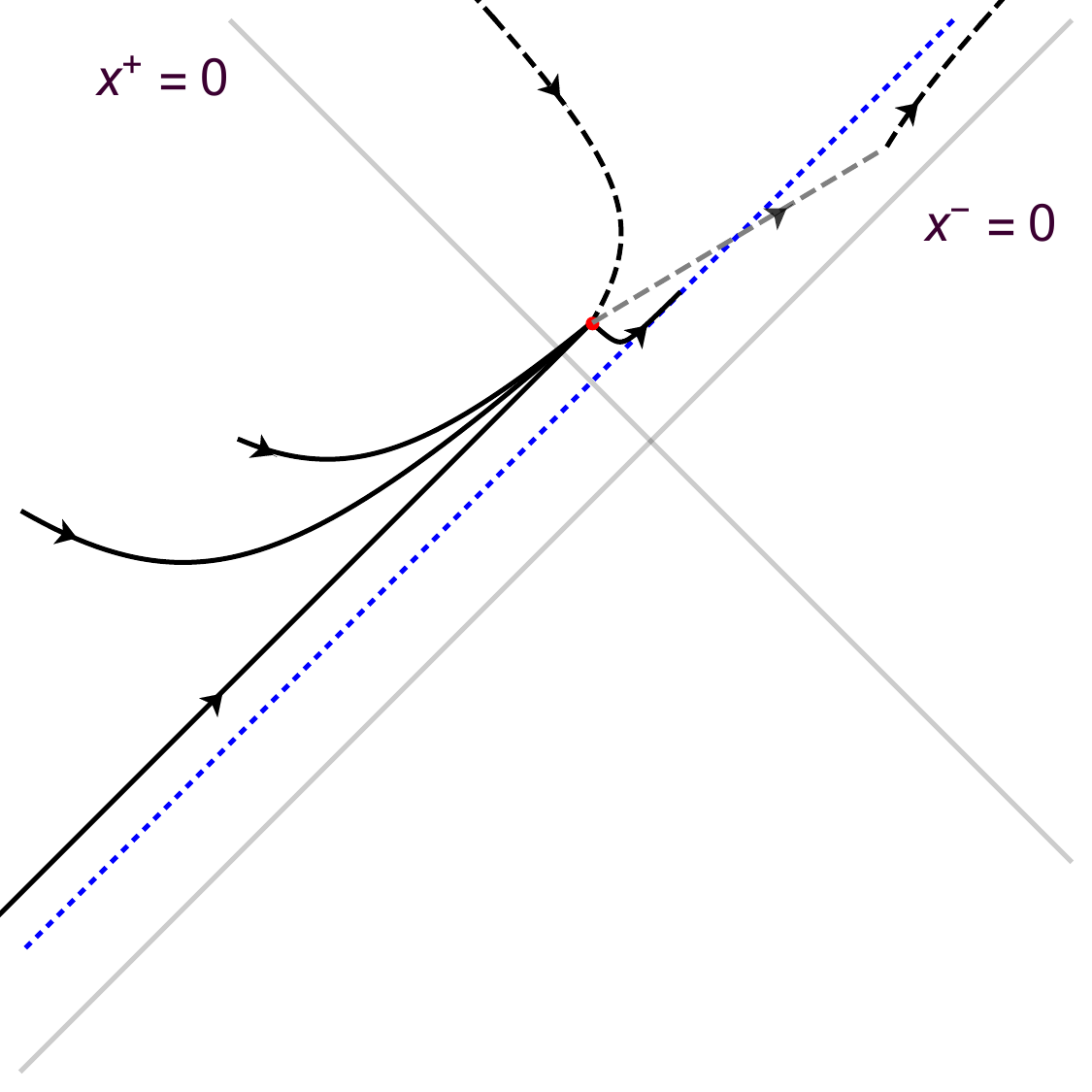}
 \caption{$x^\LCm > x^\LCm_h$}   
    \label{fig:WL_no-real-saddle}
    \end{subfigure}
%%%%%%%%%%%%%    
    \caption{\label{fig:EM-horizon} 
    Saddle point worldlines $z_{\rm cl}(\tau)$ in the $x^{-}$-$x^{+}$ plane: solid curves show four representative worldlines contributing to \eqref{integral-to-do}, for fixed $z_{\rm cl}(T)=x$ (solid red dot) and $p_{\mu}$, but different $T$. The dashed blue line is the `horizon'. The classical trajectory, which exists only in case (a), is represented by the dashed curve. The real part of a representative worldline instanton corresponding to a complex $T_s$ is shown as the dashed curve in (b).}
\end{figure}

The integral \eqref{integral-to-do} is still a sum over worldlines, but now  over the \emph{saddle point} worldlines $z_{\rm cl}(\tau)$, which we can think of as parametrised by $T$. These are not the classical trajectories discussed in (\ref{subsec:classical_EM}), as they can be both timelike and spacelike, as seen by considering
\begin{align}\label{off-shell}
    \frac{1}{4}\dot{z}_{\rm cl}^2    =
    2p_{\LCm}p_{\LCp}
    \bigg(1-\frac{x^\LCm}{x^\LCm_h}\bigg)e^{2eET}-p_{\LCperp}^2\;.
\end{align}
In the classically allowed regime, $x^\LCm<x^\LCm_h$, the right hand side of (\ref{off-shell}) can be positive or negative. In the classically forbidden region $x^\LCm>x^\LCm_h$, though, contributions to (\ref{integral-to-do}), thus to the wavefunction $\varphi_{p}(x)$, come exclusively from spacelike worldlines, see Fig~\ref{fig:EM-horizon}. This suggests that the two cases $x^\LCm\gtrless x^\LCm_h$ will need to be treated separately, as is immediately born out when we identify the saddle points $T_s$ of (\ref{integral-to-do}). These obey $S_{\rm cl}'(T_{s})=0$, or 
\begin{align}\label{saddle-cond-1}
    e^{-2eET_{s}} = 1-\frac{x^{\LCm}}{x^\LCm_h} \;,
\end{align}
(This saddle point condition is  equivalent to $\dot{z}_{\rm cl}^2/4-m^2=0$ for classical worldlines.) Outside the horizon, $x^\LCm < x^\LCm_h$, the right hand side of (\ref{saddle-cond-1}) is positive and the saddle points lie at
\be\label{saddles-outside}
    T_{s} = -\frac{1}{2eE}\log\bigg(1-\frac{x^{\LCm}}{x^\LCm_h}\bigg) - \frac{n \pi i}{eE} \;, \quad n \in \mathbb{Z} \;,
\ee
in particular the $n=0$ saddle lies on the real axis. Beyond the horizon, on the other hand, $x^\LCm>x^\LCm_h$, the right hand side of (\ref{saddle-cond-1}) becomes negative and there is no real saddle. Instead all saddle points lie off the real axis at
\be\label{complex-saddles}
    T_s = \frac{-i \pi}{2eE} -\frac{1}{2eE} \log  \bigg(\frac{x^{\LCm}}{x^\LCm_h}-1\bigg) - \frac{n\pi i}{eE}  \;, \quad n \in\mathbb{Z}  \;.
\ee
It is useful to now look at the explicit form of the corresponding saddle point trajectories. 

\begin{figure}[t!]
    \centering
    \includegraphics[width=.5\textwidth]{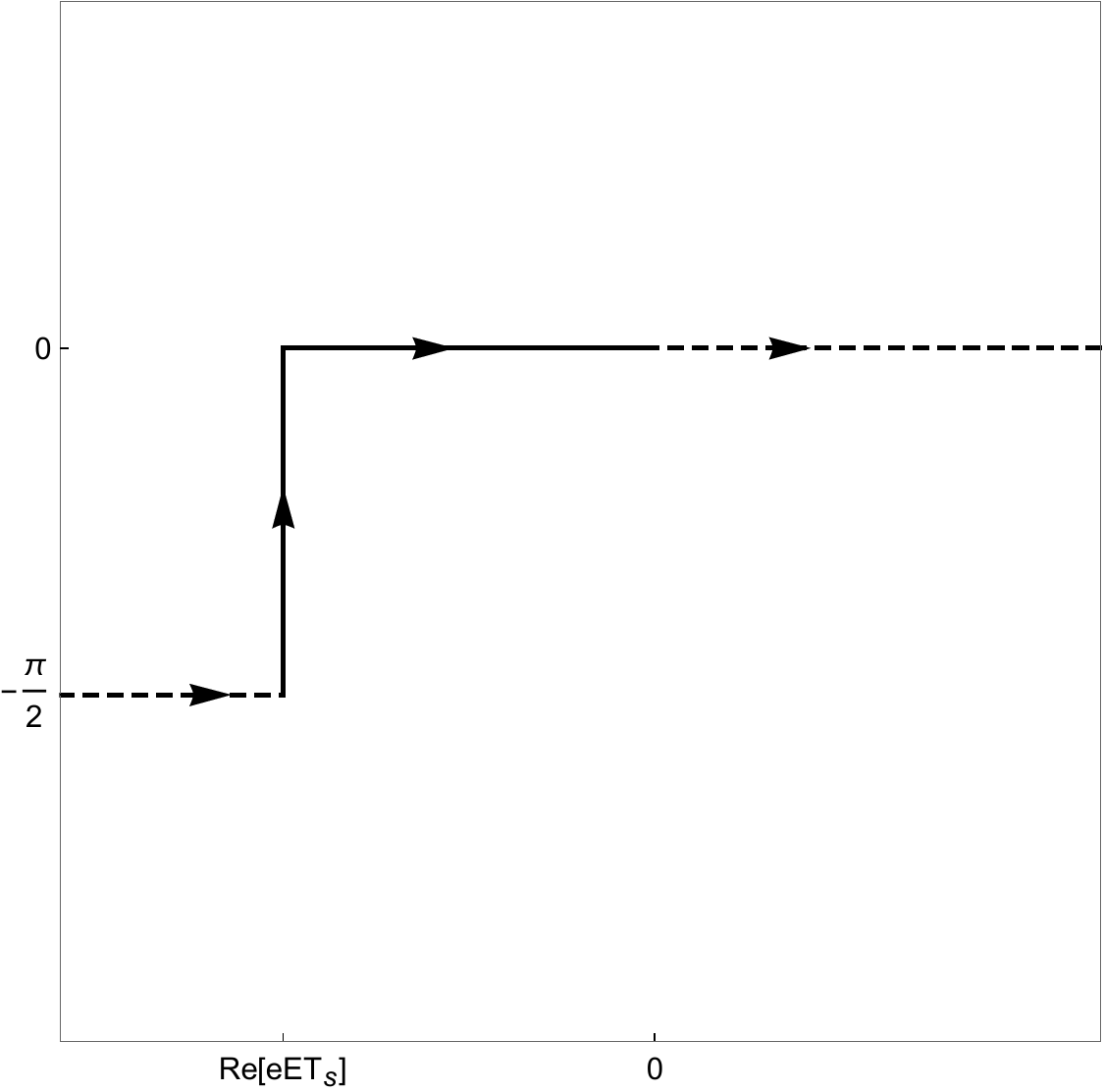}
    \caption{\label{fig:propertime}
        A representative contour in the complex $\tau$-plane for the instanton describing the tunnelling wavefunction beyond the horizon ($x^{\LCm}>x_{h}^{\LCm}$). The thick curves cover the finite range relevant to computation of the wavefunction, while the dashed curves extend the range of proper time such that the instanton corresponds to the (dashed) worldline in Fig.~\ref{fig:WL_no-real-saddle}. }
\end{figure}

Outside the horizon, substitution of the saddle point condition \eqref{saddles-outside} into \eqref{eq:z_cl_tau} yields a classical worldline (i.e.~a solution of the Lorentz force equation) with the form
\begin{align}\label{eq:z_classical_expl.}
    z^{\LCm}_{\rm cl}(\tau)\Big\rvert_{T\rightarrow T_s}=x^\LCm_h(1-e^{-2eE\tau})\,,
\end{align}
where we focus on the minus-component. This worldline always lies in the classically allowed region $z_{\rm cl}^\LCm < x^\LCm_h$, even if $\tau$ varies over the whole real line, see Fig.~\ref{fig:WL_real-saddle}.
%%%%
For $x^\LCm> x^\LCm_h$ on the other hand, the saddle point value of $T_{s}$ is complex, implying that $z^{\LCm}_{\rm cl}(\tau)$ in \eqref{eq:z_classical_expl.} can be complex, as $\tau$ itself is now allowed to take complex values.
The worldlines are thus only defined up to contour deformations. Nevertheless, some physical insight can still be obtained by choosing a representative contour in the complex $\tau$-plane, for example that shown in Fig.~\ref{fig:propertime}.
The saddle point worldline is, explicitly, writing $r \equiv \text{Re}[T_s]\in\mathbb{R}$,
\begin{align}\label{eq:z_classical_expl_2}
    z^{\LCm}_{\rm cl}(\tau)|_{T\to T_s} = \begin{cases}
        x^\LCm_h(1-e^{-2eE t})\;, & \tau = t \in [r,\infty)\\[2pt]
        x^\LCm_h(1-e^{-i\pi s}e^{-2eE\,r}) & \tau = \displaystyle\frac{i \pi}{2eE} s+r\;, \quad s \in(-1,0)\\[5pt]
        x^\LCm_h(1+e^{-2eE t}) & \tau = \displaystyle\frac{-i\pi}{2eE}+t \;, \quad t \in (-\infty,r] \;.
    \end{cases}
\end{align}
When $\tau$ lies on the positive real axis, see Fig.~\ref{fig:propertime}, we recover precisely \eqref{eq:z_classical_expl.}, which is again a particle worldline lying outside the horizon. Now, when $\tau$ lies on the lower horizontal line in Fig.~\ref{fig:propertime}, the trajectory remains real and describes an \emph{antiparticle} worldline, parametrized by real $\tau_r$, which always lies \emph{beyond} the horizon. Therefore, the complex worldline describing tunneling can be thought of the union of real particle and anti-particle trajectories bridged by a complex worldline, as shown in Fig.~\ref{fig:WL_no-real-saddle}. These serve as proxies for correlations associated with the spacelike worldlines on which the path integral has support when $x^\LCm>x^\LCm_h$. 

These saddle point solutions, or `worldline instantons', are commonly encountered in the Schwinger effect -- 
closed instantons are related to the imaginary part of the effective action and the integrated pair creation probability~\cite{AFFLECK1982509,Kim:2000un,Dunne:2005sx,Dunne:2006st,Dunne:2006ur,Dumlu:2011cc,Ilderton:2015qda,Dumlu:2017kfp}, while differential rates and spectra can be accessed using open worldline instantons~\cite{DegliEsposti:2021its,DegliEsposti:2022yqw,DegliEsposti:2023qqu,DegliEsposti:2024upq}. Ours are examples of the latter.
%

%%%%%%%%%%%%%%%%%%%%%%%%%%
%%%%%%%%%%%%%%%%%%%%

\begin{figure}[t!!]
    \centering
%%%%%%%%%%%%%%%   
    \begin{subfigure}[t]{0.45\textwidth}
    \includegraphics[width=\textwidth]{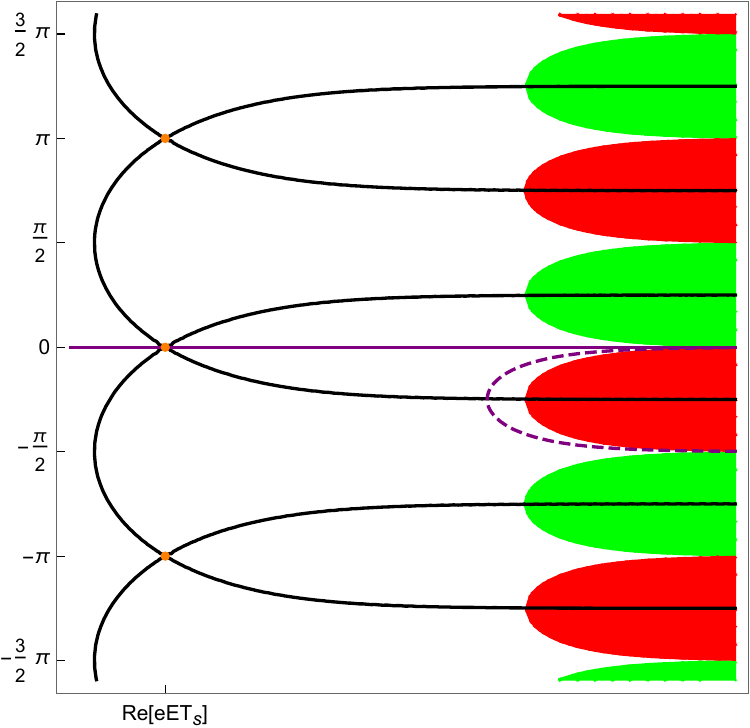}
\caption{$x^\LCm < x^\LCm_h$, a real saddle point exists.}
    \label{fig:real-saddle}
    \end{subfigure}
%%%%%%%%%%%%%%
     \hfill
 %%%%%%%%%%%%%    
    \begin{subfigure}[t]{0.45\textwidth}
    \includegraphics[width=\textwidth]{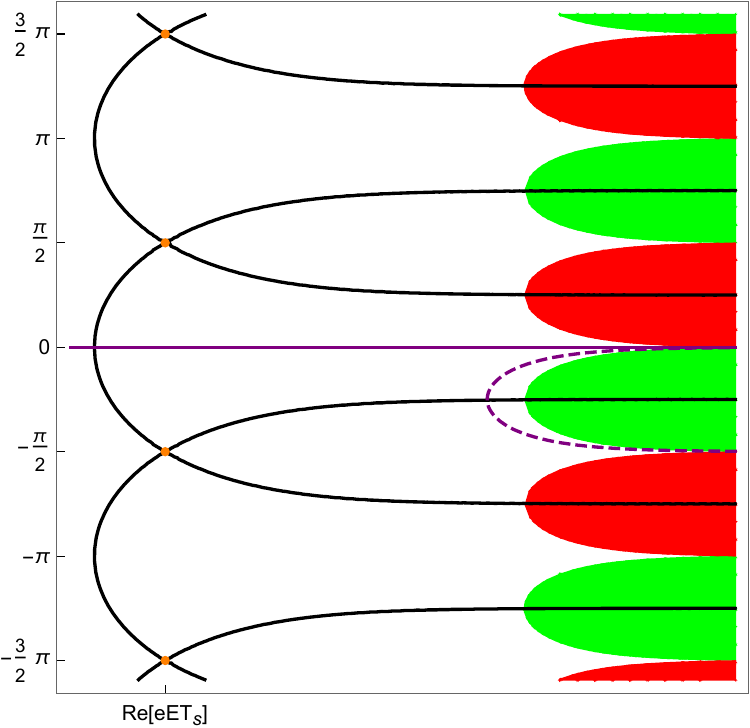}
 \caption{$x^\LCm > x^\LCm_h$, a real saddle point exists.}   
    \label{fig:no-real-saddle}
    \end{subfigure}
%%%%%%%%%%%%%    
    \caption{
      \label{fig:S-in-complex-T}
      $\mathcal{S}_{\rm cl}(T)$ in complex $(eET)$-plane. Black curves: steepest descent/ascent curves intersecting at the saddle points(Orange), Purple continuous: original contour along real line, Purple dashed: contour to get non-Tunnelling mode function,   Red region: $\textrm{Re}[\mathcal{S}_{\rm cl}]$ is very large positive, Green region: $\textrm{Re}[\mathcal{S}_{\rm cl}]$ is very large negative.}
\end{figure}

We now evaluate the $T$-integral. We could do this by integrating along the steepest descent contours, which are shown along with the saddle points in Fig.~\ref{fig:real-saddle} for $x^\LCm < x^\LCm_h$ and in Fig.~\ref{fig:no-real-saddle} for $x^\LCm > x^\LCm_h$. The steepest descent contours pick up contributions from all saddles, as we discuss later, but here we present a more direct route to the wavefunction of interest.  

For $x^\LCm < x^\LCm_h$ we shift the integration variable by the real ($n=0$) saddle, $T \to T+T_s$. This pulls all $x^\mu$-dependence out of the proper time integral which thus  contributes a normalisation constant, $\mathcal{N}$. The integral is convergent, since we can simply displace the contour into the upper half plane, such that it runs to the green-shaded region of Fig.~\ref{fig:real-saddle} just above the real axis. For $x^\LCm>x^\LCm_h$, the singularities of $iS_\text{cl}$ tell us, see Fig.~\ref{fig:no-real-saddle}, that we can displace our contour to run through the first complex saddle \emph{below} the real axis, i.e.~$n=0$ in (\ref{complex-saddles}). Again, this pulls all $x^\mu$-dependence out of the proper-time integral. One finds, combining both cases,
\begin{align}\label{stack}
    \varphi_{p}(x) &=
    \mathcal{N}
    e^{-ip_\LCp x^\LCp -ip_\LCperp x^\LCperp}
    \begin{cases}
        e^{-\big(\frac{1}{2}-i \frac{m^2+p_{\perp}^2}{2eE}\big)\log\left(1-\frac{x^\LCm}{x^\LCm_h }\right)} & x^\LCm < x^\LCm_h \;, \\
        {e^{-\pi\frac{p_\LCperp^2+m^2}{2eE}}}
    e^{-i\frac{\pi}{2}}e^{-\big(\frac{1}{2}-i \frac{m^2+p_{\perp}^2}{2eE}\big)\log\left(\frac{x^\LCm}{x^\LCm_h }-1\right)} & x^\LCm > x^\LCm_h \;,
    \end{cases}
\end{align}
Thus, outside the horizon, we recover (\ref{eq:incoming-basic}). 
Beyond the horizon, we immediately recognise the expected tunnelling exponential, which is non-perturbative in the field coupling $eE$. The sign of the exponent (damping rather than an unphysical exponential increase) is fixed by the allowed contour deformations. The quantum wavefunction thus tunnels into the classically forbidden region, where it is exponentially damped. 

%
%%%%%%%%%%%%%%%%%%%%%%%%%%
%%%%%%%%%%%%%%%%%%%%%%%%%

\begin{figure}[t!!]
    \centering
%%%%%%%%%%%%%%%   
    \begin{subfigure}[t]{0.45\textwidth}
    \includegraphics[width=\textwidth]{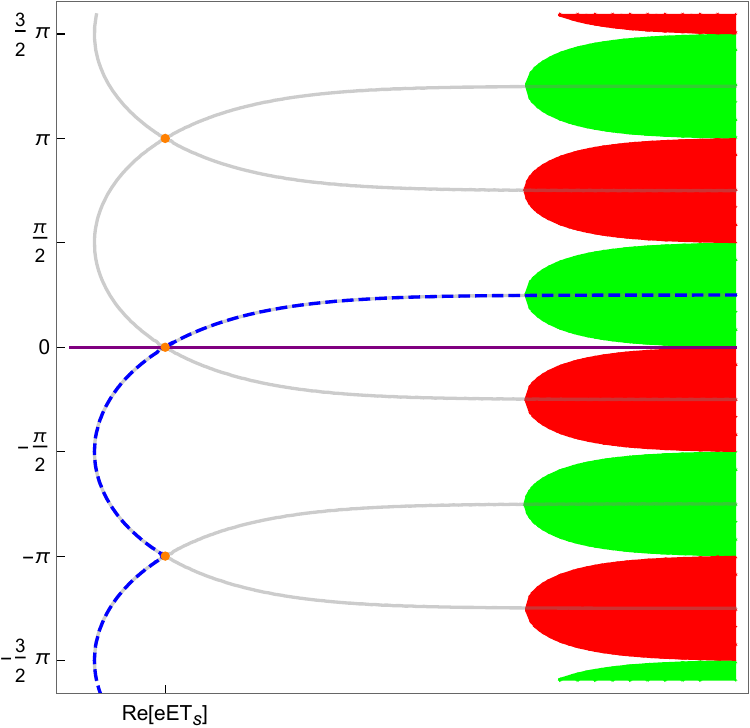}
    \caption{$x^\LCm < x^\LCm_h$, a real saddle point exists.}
    \label{fig:real-saddle-deform}
    \end{subfigure}
%%%%%%%%%%%%%%
     \hfill
 %%%%%%%%%%%%%    
    \begin{subfigure}[t]{0.45\textwidth}
    \includegraphics[width=\textwidth]{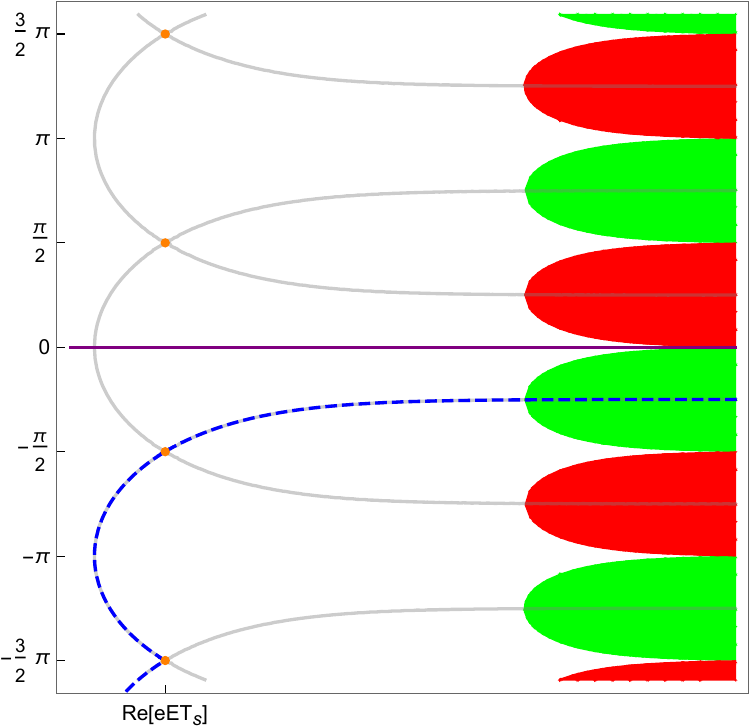}
    \caption{$x^\LCm > x^\LCm_h$, a real saddle point does not exist.}
    \label{fig:no-real-saddle-deform}
    \end{subfigure}
%%%%%%%%%%%%%    
    \caption{The steepest descent contours (dashed-blue) for evaluating the $T-$integral pass through all saddle points below the real line, both outside and beyond the horizon.} \label{fig:S-in-complex-T-deform}
\end{figure}

The normalisation constant is
\be\label{discard1}
\begin{split}
    \mathcal{N} = \int_{-\infty}^{\infty}\!\ud T\, e^{eE T} \exp\left[i(e^{2eE{T}}-1)\frac{p_\LCperp^2 +m^2}{2eE}-i(p_{\perp}^2+m^2){T}\right]  \;.
\end{split}
\ee
The integral can be performed in terms of the Gamma function, but we will only need its modulus. Any other ambiguities can be fixed by the limit $eE\to 0$. We can ultimately write
\be
    \mathcal{N} = \frac{1}{\sqrt{1+e^{-\pi (p_\LCperp^2+m^2)/(eE)}}} \;.
\ee
We may write a single expression for $\varphi_p(x)$ valid for all $x^\LCm$ by noting that the functional form in (\ref{stack}) corresponds to using the principal value of log. With this understood, we have
\be\label{full-tunnelling-phi}
    \varphi_{p}(x) =
    \mathcal{N}\, e^{-ip_\LCp x^\LCp -ip_\LCperp x^\LCperp}
    e^{-\big(\frac{1}{2}-i \frac{m^2+p_{\perp}^2}{2eE}\big)\log
    \left(1-\frac{x^\LCm}{x^\LCm_h }\right)}\;,
    \quad \forall\, x^\LCm\;.
    \ee
The classical physics of this system suggests that the the 1-to-1  scattering amplitude should be encoded in the behaviour of $\varphi_p(x)$ for $x^\LCm<x^\LCm_h$, while our quantum expectations are that the pair creation amplitude should be encoded in $\varphi_p(x)$ in the classically forbidden region $x^\LCm>x^\LCm_h$. As we confirm in detail in Appendix.~\ref{app:Schwinger}, $\mathcal{N}$ is essentialy the 1-to-1 amplitude. As we cross the horizon, this is multiplied by the Schwinger factor in (\ref{stack}), and this is the pair creation amplitude. In other words we can read off from $\varphi_p(x)$ the (mod-squared) amplitudes
\be
 |\mathcal{M}_{1\to1}|^2 = \frac{1}{1+e^{-\lambda}} \;,
        \qquad
        |\mathcal{M}_{0\to2}|^2 = \frac{e^{-\lambda}}{1+e^{-\lambda}} \;,
        \qquad
        \lambda = \frac{p_\LCperp^2+m^2}{eE} \;.
    \ee
Before moving on, we clarify the role of contributions from the infinite tower of saddle points in (\ref{saddles-outside}) and (\ref{complex-saddles}). In principle, \textit{all} saddles below the real axis contribute to the wavefunction. This becomes evident by deforming the original $T$-integral contour so that it passes through all {relevant} saddles, see Fig.~\ref{fig:S-in-complex-T-deform}. The resulting integral is (unlike~\eqref{discard1}) \emph{manifestly} convergent. This implies a tunnelling-like effect even within the classically allowed region -- these saddles correspond to complex worldlines that cross the horizon multiple times.

%%%%%%%%%%%%%%%%%%
%%%%%%%%%%%%%%%%%%%%%%%%%%%

%%%%%%%%%%%%
\subsubsection{Properties of tunnelling and other modes}
%%%%%%%%%%%%
More properties of the tunnelling wavefunction (\ref{full-tunnelling-phi}) are highlighted by comparison with other possible solutions of the Klein-Gordon equation. These can be obtained by, for example, choosing a different contour for the proper time integral. The purple dashed curve in Fig.\ref{fig:S-in-complex-T} is one such choice; for $x<x^\LCm_h$ the contour integral reproduces the same function (\ref{eq:incoming-basic}) outside the horizon, as above, but for $x>x^\LCm_h$ the contour can be contracted to a point, and the integral vanishes. One therefore obtains 
\begin{align}\label{classical-like-solution}
    \int_{\infty}^{\infty-i\frac{\pi}{2eE}}\ud T\!\int^{z(T)=x}\!\mathcal{D}[z]\,
    e^{i\mathcal{S}_{\rm WL}}
    \propto  \Theta(x^\LCm_h  -x^\LCm)\, \varphi_p(x)=:\phi_p(x) \;.
\end{align}
The contour in (\ref{classical-like-solution}) thus gives a solution which is somewhat more representative of the classical discussion in Sec.~\ref{subsec:classical_EM} -- the wavefunction vanishes beyond the horizon that a classical particle cannot cross. (It is straightfoward to check that despite the hard cutoff in the wavefunction, it obeys the Klein-Gordon equation everywhere.) Note there are no contributions from multiple saddles for $\phi_p$, hence the normalisation differs from that of the tunnelling wavefunction (which we revisit in the next section), and so all contributing saddle point worldlines lie outside the horizon.

It looks from (\ref{eq:KG-simplified}) that, in the chosen gauge, boundary conditions are naturally placed on surfaces of constant $x^\LCm$.
This provides data on a characteristic, rather than Cauchy data -- see ~\cite{Tomaras:2000ag,Tomaras:2001vs} for a comprehensive discussion.
The choice of proper-time contour seems to be a way to impose extra conditions which leads to different solutions.  The lack of good asymptotics obscures this, but if we imagine that the field is turned off at some finite $x^\LCp$ (note), then within the field our solutions are still given by the expressions above. Fourier transforming the fields on the constant $x^\LCp$ surface allows us to inspect their spectral composition.

The relevant integrals are easily performed in terms of the gamma function. Writing out only the nontrivial structures, we find
\be\label{Fourier1}
  \int\limits_{-\infty}^{\infty}\!\ud x^\LCm e^{i\omega x^\LCm} \phi_p(x) \sim (i \omega x_h^\LCm)^{-i \frac{\lambda}{2} -\frac12}x_h^\LCm e^{i\omega x^\LCm_h} \Gamma\big(i \frac{\lambda}{2} +\tfrac12\big) \;,
\ee
which in particular has support on all frequencies $\omega$.  Both the explicit step function in $\phi_{p}(x)$, and its Fourier transform, resemble structures in the `out'-modes relevant to Hawking radiation~\cite{Hawking:1975vcx}, which we discuss in Sec.~\ref{sec:hawking}. The Fourier transform of the tunnelling solution, on the other hand, is 
\be\label{Fourier2}
    \int\limits_{-\infty}^{\infty}\!\ud x^\LCm e^{i\omega x^\LCm} \varphi_p(x) \sim (i \omega x_h^\LCm)^{-i \frac{\lambda}{2} -\frac12}x_h^\LCm e^{i\omega x^\LCm_h} \Gamma\big(i\dfrac{\lambda}{2}   +\tfrac12\big)\big(1+{e^{-\pi \lambda}}\big)\Theta(\omega) \;,
\ee
which, in contrast to (\ref{Fourier1}) has support only on $\omega>0$. In this sense, the tunnelling solution $\varphi_p$, which has support both outside and beyond the horizon, is a positive energy mode.

\section{Amplitudes from mode functions in background fields}\label{sec:modes_amplitudes}
Particle creation in a background, be it in gauge theory or gravity, is often described in terms of Bogoliubov coefficients relating incoming and outgoing modes of the corresponding field operator (assuming a quadratic Hamiltonian, as will be the case throughout). The ultimate objects of interest, though, are asymptotic observables, which should be expressible in terms of amplitudes. The purpose of this section is therefore to give some general results linking the Bogoliubov and amplitudes-based approaches, in any gauge or gravitational background with sufficiently good behaviour to admit a notion of asymptotic states. (For a recent discussion of the subtleties concerning asymptotic states when long-range forces are involved see~\cite{Lippstreu:2025jit}.) Combined with intuition from the Schwinger effect, this will set the stage for our later discussion of Hawking radiation.

Our starting point is the appropriate scalar field equation -- the Klein-Gordon equation in QED or wave equation in gravity -- and its `in' and `out' mode solutions.
These reduce to free solutions in the asymptotic past and future (as appropriately defined for massive or massless particles) respectively. For in, out and free wavefunctions parameterised by a set of quantum numbers `$\ii$' we write $\varphi_\ii^{\pm\text{in}}$, $\varphi_\ii^{\pm\text{out}}$, $\varphi_\ii^{\pm\text{free}}$ respectively, in which `$\pm$' refers to positive and negative energy modes (particles and antiparticles in QED). The scalar field operator $\Phi(x)$ can be expanded using either the in or out modes as 
\begin{align}\label{in-and-out-expand}
    \Phi(x)&=
    b_\ii \varphi^{+\text{in}}_\ii(x) + d_\ii^\dagger \varphi^{-\text{in}}_\ii (x) 
     = B_\ii \varphi^{+\text{out}}_\ii(x) + D_\ii^\dagger \varphi^{-\text{out}}_\ii (x) \;,
\end{align}
in which an appropriate sum/integral over repeated indices is implied. The set of `in' operators defines the in-vacuum via $b_\ii \ket{\rm in}=d_\ii \ket{\rm in}=0$ and, similarly, the out-vacuum\footnote{The `in' vacuum is naturally taken to be the usual, empty vacuum in which our initial scattering states are prepared. The `out' vacuum is just some other reference state, \emph{not} the time-evolved `in' vacuum.} is defined by $B_\ii \ket{\rm out}=D_\ii \ket{\rm out}=0$.

Linearity of the wave equation implies that in-modes can be expressed as linear combinations of out-modes, and vice versa. Equivalently, the form of the Hamiltonian implies that the in-operators in (\ref{in-and-out-expand}) are linear combinations of the out-operators. We may write
\begin{align}
    \begin{pmatrix}
        b_\ii \\ d^\dagger_\ii 
    \end{pmatrix}
    =
    \begin{pmatrix}
        \alpha_{\ii\,\jj} & \beta_{\ii\,\jj} \\[3pt]
        \beta^\dagger_{\ii\,\jj} & \alpha^\dagger_{\ii\,\jj}  
    \end{pmatrix}
    \begin{pmatrix}
        B_\jj \\ D^\dagger_\jj 
    \end{pmatrix} \;,
\end{align}
in which we adopt a matrix notation for the \emph{Bogoliubov coefficients} such that, e.g.~$\alpha^\dagger_{\ii\,\jj} = \alpha^\star_{\jj\,\ii}$. The field operator in the asymptotic future is related to that in the past by the $S$-matrix, hence $B_\ii =S^{\dagger}b_\ii S$ and so on, while unitarity of the transformation is expressed as usual by
\be
    (\alpha\alpha^\dagger)_{\ii\,\jj} - (\beta\beta^\dagger)_{\ii\,\jj} = \delta_{\ii\,\jj} \;,
    \qquad
    (\alpha\beta)_{\ii\,\jj} - (\beta\alpha)_{\ii\,\jj} = 0 \;.
\ee
The in and out modes themselves are related by
\begin{align}
\label{eq:outmode_gen}
\begin{pmatrix}
    \phi^{+\rm in}_\ii & \phi^{-\rm in}_\ii
\end{pmatrix}
=
\begin{pmatrix}
    \phi^{+\rm out}_\jj & \phi^{-\rm out}_\jj
\end{pmatrix}
 \begin{pmatrix}
        \alpha^\dagger_{\jj\,\ii} & -\beta_{\jj\,\ii} \\[3pt]
        -\beta^\dagger_{\jj\,\ii} & \alpha_{\jj\,\ii}
    \end{pmatrix} \;.
\end{align}
To illustrate the information content of the Bogoliubov coefficients, we can compute e.g.~the expected number of particles in the asymptotic future, starting from the empty vacuum $\ket{\text{in}}$:
\begin{align}\label{eq:particle_num_gen}
    \braket{{\rm in}|S^{\dagger} {b^\dagger}_{\!\ii} b_\ii S|{\rm in}}=
    \beta^\dagger_{\ii\,\jj}{\beta}_{\jj\,\ii}\;, \quad\text{(no sum over $\ii$)}\;.
\end{align}
This expression and equations \eqref{eq:outmode_gen} show that the out-modes evaluated at \textit{past} asymptotic infinity encapsulate, via the Bogoliubov coefficients, information about particle creation at \textit{future} asymptotic infinity. The Bogoluibov coefficients must also encode the amplitudes of the background QFT, namely the $1\rightarrow 1$ scattering and $0\to2$ pair creation amplitudes,
\begin{align}
    \braket{{\rm in}|{b_\jj} S {b^\dagger}_{\!\ii}|{\rm in}} \;,
    \qquad
    \braket{{\rm in}|{d_\jj} S {d^\dagger}_{\!\ii}|{\rm in}} \;,
    \qquad
    \braket{{\rm in}| d_\jj b_\ii S|{\rm in}}\,.
\end{align}
To connect these amplitudes to the Bogoliubov coefficients, it is convenient to divide through by the (non-trivial) vacuum persistence amplitude $\braket{{\rm in}|S|{\rm in}}$ which appears as a factor in all amplitudes; this defines the corresponding \emph{diagrams}, see Fig.~\ref{fig:amps-to-diagrams}. It is these diagrams which can be written explicitly in terms of the Bogoliubov coefficients as~\cite{Fradkin1991QED,Wald1994QFT}:
\begin{align}
\label{amp:one-to-one}
    \mathcal{A}_{1\rightarrow 1}(\ii\to \jj)&\equiv\frac{ \braket{{\rm in}|S^{\dagger} {b^\dagger}_{\!\jj} b_\ii S|{\rm in}}}{\braket{{\rm in}|S|{\rm in}}}=(\alpha^{-1})_{\jj\,\ii} \;,\\
\label{amp:zero-to-two}
\mathcal{A}_{0\rightarrow 2}(\mathbf{i},\mathbf{j})&\equiv\frac{\braket{{\rm in}| d_\jj b_\ii S|{\rm in}}}{\braket{{\rm in}|S|{\rm in}}}=-\left(\beta \alpha^{-1}\right)_{\ii\,\jj} \;.
\end{align}

%%%%%%%%%%%%%%%%%%%%%%%%%%
\begin{figure}[t!]
\centering
\begin{align*}
\braket{{\rm in}|S|{\rm in}}&\quad=\quad \exp\left(
    \raisebox{1pt}{\begin{tikzpicture}[baseline=-2pt]
     \draw[double] (0,0) circle (.2);%bubble
    \end{tikzpicture}}
    \right) \\[-5pt]
    %%%%%%%%%%%%%%%%%%%%%%%%%%%%%%%%%%%%%%%%%%%%%%%%%%%%%
       \braket{{\rm in}|{b_\jj} S {b^\dagger}_{\!\ii}|{\rm in}}&\quad=\quad  \begin{tikzpicture}[baseline=-2pt]
    \draw[double] (-1,0) -- (.3,0);%line
    \end{tikzpicture}\;+\;
    \begin{tikzpicture}[baseline=-2pt]
    \draw[double] (-1,0) -- (.3,0);%line
    \draw[double] (-.4,.6) circle (.2);%bubble
    \end{tikzpicture}\; +\;\frac{1}{2!}
    \begin{tikzpicture}[baseline=-2pt]
    \draw[double] (-1,0) -- (.3,0);%line
    \draw[double] (-.5,.6) circle (.2);%bubble
     \draw[double] (0,.6) circle (.2);%bubble
    \end{tikzpicture}\;+\ldots
    = \exp\left(
    \raisebox{1pt}{\begin{tikzpicture}[baseline=-2pt]
     \draw[double] (0,0) circle (.2);%bubble
    \end{tikzpicture}}
    \right) \begin{tikzpicture}[baseline=-2pt]
    \draw[double] (-1,0) -- (.3,0);%line
    \end{tikzpicture}
    \\[5pt]
    %%%%%%%%%%%%%%%%%%%%%%%%%%%%%%%%%%%%%%%%%%%%%%%%%%%%%%%
   \braket{{\rm in}|d_\jj b_\ii S|{\rm in}}&\quad=\quad
   \exp\left(
    \raisebox{1pt}{\begin{tikzpicture}[baseline=-2pt]
     \draw[double] (0,0) circle (.2);%bubble
    \end{tikzpicture}}
    \right) \begin{tikzpicture}[baseline=-2pt]
    \draw[double] (0,.3) -- (.3,.3);%line_of_horseshoe
    \draw[double] (0,-.3) -- (.3,-.3);%line_of_horseshoe
    \draw[double] (0,0.3) arc (90:270:.3);%arc_of_horseshoe
    \end{tikzpicture}
\end{align*}
%%%%%%%%%%%%%%%%%
    \caption{\label{fig:amps-to-diagrams} Top to bottom: the vacuum persistence, 1-to-1 and pair creation amplitudes. Double lines indicate (suitably amputated) background-field propagators, in which the coupling to the background is treated exactly. Bubble diagrams exponentiate in every amplitude; the exponent has a non-trivial real part which contributes to physical quantities. Dividing out the exponential (only) for convenience, what remains are the \emph{diagrams} one would naturally write down from the Feynman rules of the theory. See~\cite{Copinger:2024pai,Aoude:2024sve} for recent discussions.}
\end{figure}

In/out modes are naturally associated with retarded/advanced boundary conditions and propagators, whereas amplitudes are naturally associated with Feynman propagators. This prompts us to consider a different set of solutions of the wave equations -- the Feynman modes. We will \emph{define} these in terms of the in-modes as
\begin{align}
\varphi^{+\text{F}}_{\ii}(x):= 
    \varphi^{+\text{in}}_{\jj}(x)({\alpha^{\dagger-1}})_{\jj\,\ii}
\qquad
   \varphi^{-\text{F}}_{\ii}(x):= 
   \varphi^{-\text{in}}_{\jj}(x)(\alpha^{-1})_{\jj\,\ii}\, \;.
\end{align}
By construction, the $\varphi^{\pm\text{F}}_{\mathbf{i}}(x)$ are positive and negative energy modes in the asymptotic past, like the in-modes. The difference is that, while the in/out-modes directly  encode the Bogoliubov coefficients, the Feynman modes directly encode the scattering and pair production amplitudes $\mathcal{A}_{1\rightarrow 1}(\mathbf{i}\rightarrow\mathbf{j})$ and $\mathcal{A}_{0\rightarrow 2}(\mathbf{i},\mathbf{j})$ through their asymptotic behaviour. Specifically, it follows from \eqref{eq:outmode_gen} and the expressions (\ref{amp:one-to-one})--(\ref{amp:zero-to-two}) that
\begin{align}\label{eq:def_Feynman_modes}
  \varphi^{-\text{F}}_{\mathbf{i}}(x)\sim\begin{cases}
      \sum_{\mathbf{j}}\mathcal{A}_{1\rightarrow 1}(\mathbf{i}\rightarrow\mathbf{j})\varphi^{-\text{free}}_{\mathbf{j}}(x)\quad&;\quad x^0\rightarrow -\infty\,,\\
      \varphi^{-\text{free}}_{\mathbf{i}}(x)+\sum_{\mathbf{j}}\mathcal{A}_{0\rightarrow 2}(\mathbf{i},\mathbf{j})\varphi^{+\text{free}}_{\mathbf{j}}(x) \quad&;\quad x^0\rightarrow +\infty\,.
  \end{cases}  
\end{align}
Amplitudes are thus encoded in the \emph{asymptotic behaviour} of the Feynman modes. 

While we defined the Feynman modes from in-modes, one could alternatively obtain them by solving the appropriate wave equation with boundary conditions implied by \eqref{eq:def_Feynman_modes}. These are that $\varphi^{-\text{F}}_{\mathbf{i}}(x)$ (1) approaches a negative energy function in the past (the first line of (\ref{eq:def_Feynman_modes}) is a sum over free, on-shell negative energy modes) and (2) is a mixture of positive and negative modes in the future, with the negative energy part being a free `$\mathbf{i}$' mode. Practically, in a perturbative solution of the wave equation in powers of the coupling, the Feynman modes follow from iterating with the free anti-/Feynman propagator~\cite{ToAppear}.

We comment that in perturbative, amplitude-based approaches, the exponentiation of amplitudes -- obtained by resumming diagrams at all orders but within a fixed order of the semiclassical expansion -- occurs after performing an on-shell Fourier transform from momentum space to spacetime. The asymptotic behaviour of the Feynman modes, as we have just seen, corresponds precisely to this transform. This makes studying the Feynman modes a natural alternative to analysing amplitudes directly, in particular in semiclassical settings, such as Hawking radiation.

\section{Hawking radiation from worldline QFT}\label{sec:hawking}
\begin{figure}[t!]
    \centering
    \includegraphics[width=0.3\linewidth]{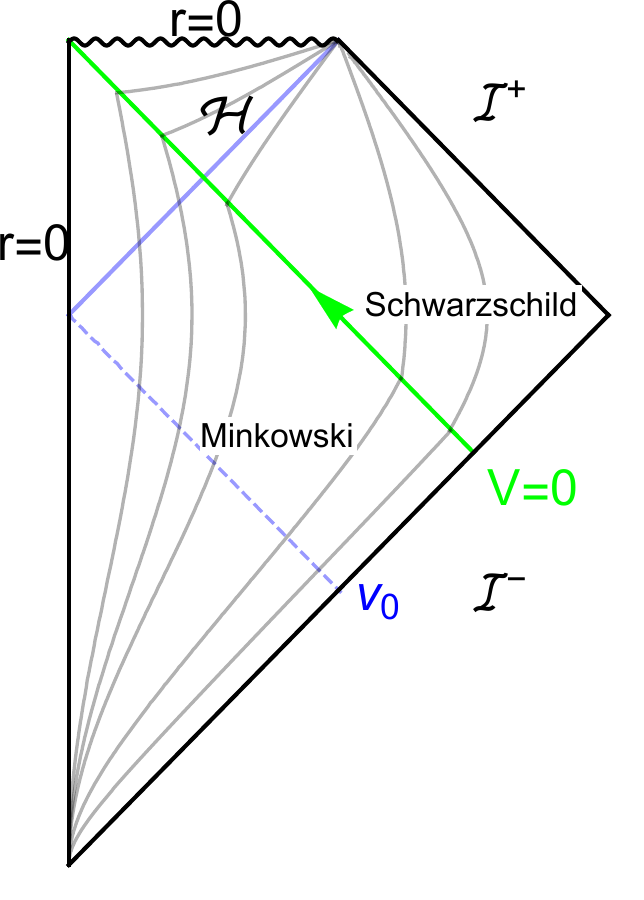}
    \caption{\label{fig:penrose_vaidya} The Penrose diagram for the Vaidya metric (\ref{eq:line-element}). The green line indicates the collapse of the null shell, so that $V<0$ is the Minkowski region, while $V>0$ is the Schwarzchild region. The solid blue lines marks the future event horizon $\mathcal{H}$. The blue dotted line indicates the last null ray, emitted from $V=v_0=-4GM$ at $\mathcal{I}^-$, which escapes the collapse. Grey curves are contours of constant radial coordinate $R$.}
\end{figure}

%%%%%%%%%%%%%%%%%%
We turn now to scattering and particle creation in collapsing black hole spacetimes.
Owing to the universality of the final result~\cite{Hawking:1975vcx} (see also~\cite{Unruh:2004zk}), it is sufficient to focus on the special case of a Vaidya spacetime describing the formation of a black hole from radial collapse of an infinitesimally thin spherical null shell. The metric is\footnote{We caution that $V$ (similarly $U$) is often reserved in the literature for the ingoing (outgoing) Kruskal-Szekeres null coordinate, but here we use $V$ ($U$) to denote the ingoing (outgoing) Eddington–Finkelstein coordinate.} 
\begin{align}\label{eq:line-element}
    \ud s^2= \bigg(1-\frac{2{G}M\Theta(V)}{R}\bigg)\ud V^2-2\ud V\ud R-R^2\ud\Omega^2 \;,
\end{align}
with $\ud\Omega^2=\ud\theta^2+\sin^2\theta \ud\phi^2$ the metric on the unit 2-sphere. To parallel the discussion in Sec.~\ref{subsec:classical_EM} we begin with the classical geodesics for a massless particle in the Vaidya spacetime, the Penrose diagram for which is given in Fig.~\ref{fig:penrose_vaidya}.

\subsection{Radial geodesics in Vaidya}\label{subsec:radial_geodesics}
We restrict to \emph{radial} geodesics such that $\dot\theta=\dot\phi=0$. The remaining geodesic equations are
\begin{align}
   \frac{\ud}{\ud\tau}\bigg[\bigg(1-\frac{2GM\Theta(V)}{R}\bigg)\dot{V}-\dot{R}\bigg]+\frac{GM\delta(V)}{R}\dot{V}^2&=0 \;,\\
   \ddot{V}+\frac{GM\Theta(V)}{R^2}\dot{V}^2&=0 \;,
\end{align}
and the mass-shell condition is $(1-2GM \theta(V)/R)\dot{V}^2- 2 \dot{V}\dot{R}=0$.

Consider the radial geodesic that starts from $\mathcal{I}^{-}$ at $V=v<0$, i.e.~in the Minkowski region, within which the equations of motions can be conveniently reduced to
\be
    {\dot V}(\dot{V} - 2 {\dot R}) = 0 \;, \quad \ddot{R}=0 \;.
\ee
The solution for the radial part is, choosing a parameterisation such that $R(0)=0$,  
\be
    R(\tau) = 2\epsilon_{in}|\tau|  \;, \quad V(\tau)<0 \;,
\ee
in which $\epsilon_{in}>0$ can be interpreted as an energy, {and the factor of $2$ is for convenience.} The behaviour of $V(\tau)$ is different depending on whether the worldline is approaching the origin (`ingoing', $\dot{R}<0$, so $\tau<0$) or moving away from it (`outgoing', $\dot{R}>0$, so $\tau>0$).  On the ingoing part, see Fig.~\ref{fig:penrose_vaidya}, we set $\dot V=0$, while on the outgoing part we set $\dot{V}-2\dot{R}=0$, from which it follows that 
\begin{align}
    V(\tau)&=v \;, \quad \tau < 0 \;, \\ V(\tau)&=v+4\epsilon_{in}\tau \;, 
    \quad \tau>0\quad \& \quad V(\tau)<0 \;.
\end{align}
At $V(\tau)=0$ we pass into the Schwarzchild region, where the geodesic equations are solved~by
\begin{align}\label{V-U-geodesic}
V(\tau)&=u+2R(\tau)+{4GM\log\bigg[\frac{R(\tau)}{2GM}-1\bigg] }\;, \\\nonumber
    R(\tau)&=R_0+2\epsilon\,\tau \;,
\end{align}
in which $u$, $R_0$ and $\epsilon$ are constants of integration to be determined. Demanding continuity of the geodesic at $V=0$ generates two constraints, while a third is identified by integrating the geodesic equation across the shock at $V=0$, which imposes 
\begin{align}
  \left(1-\frac{2GM\Theta(V)}{R}\right)\dot{V}-\dot{R}\Bigg\rvert_{V=0^{-}}^{V=0^+}=-\lim_{V\rightarrow 0}\frac{GM|\dot{V}|}{R} \;,
\end{align}
thus also ${\dot V}$ should be continuous. The three constraints yield, defining $v_0 = -4 GM < 0$,
\be\label{eq:constraints_geodesic}
    {R_0 = 2GM} \;, \qquad \frac{\epsilon}{\epsilon_{in}} = \frac{v}{v-v_0} \;,
    \qquad u = v - 4GM \log\bigg[\frac{v-v_0}{v_0}\bigg] \;.
\ee
The behaviour of the solution depends on whether $v<v_0$, $v=v_0$ or $v>v_0$, where $v_0$ corresponds to the last radial null geodesics that escapes the collapse. For $v<v_0$ the solution is well-behaved; the relation between incoming and outgoing energies reflects the well-known gravitational red shift in the Schwarzschild metric, while $u$ is the  point on future null infinity $\mathcal{I}^{+}$ reached by the particle. This is seen by switching to the usual outgoing coordinates, parameterising $\mathcal{I}^{+}$ with
\be\label{U-def}
    U:= V-2R-4GM \log\bigg(\frac{R}{2GM}-1\bigg) \;, \qquad R> 2GM \;,
\ee
for then $U\to u$ as $\tau\to\infty$.
The situation is different for massless particles emanating from $v>v_0$ on $\mathcal{I}^{-}$. Here the outgoing energy $\epsilon$ changes sign, becoming negative after interaction with the gravitational field of the null shell, and the asymptotic position $u$ becomes complex. Both these results reflect of the fact that the particle becomes trapped inside the horizon.

The evolution of trajectories from positive to negative energy echoes the changes in spectral properties of the wavefunctions of Sec.~\ref{sec:modes_amplitudes} -- pair production is signalled by the appearance of e.g.~negative energy modes in what was a positive energy wavefunction. In fact the connection is more concrete. Recalling Sec.~\ref{sec:tunnel_EM}, particle wavefunctions have a worldline representation. Evaluated semiclassically, the wavefunction will be controlled by saddle points of the path integral, and these are the classical geodesics. We exploit this below.

%%%%%%%%%%%%%%%%%%%%%%
\subsection{Worldline representation of wavefunctions}\label{subsec:PI_for_Hawking}
%%%%%%%%%%
We now translate the gauge theory framework of Sec.~\ref{subsec:WL_EM} to gravity, discussing some general properties of the worldline path integrals for massless particle wavefunctions. Paralleling  Sec.~\ref{subsec:WL_EM}, we can interpret the wave operator in a curved spacetime $g_{\mu\nu}$ as a Hamiltonian $H=g^{\mu\nu}\nabla_{\mu}\nabla_{\nu}$
where $\nabla_{\mu}$ is the metric covariant derivative. We can then define a proper-time kernel analogous to that in (\ref{prop-worldline-def}) and (\ref{sol-worldline-def}) via~\cite{Parker:1979mf}
\begin{align}
\braket{x|e^{-iHT}|y}   =\int_{z(-T/2)=y}^{z(T/2)=x} \mathcal{D}[z]\exp\left[i S\right]  \;,
\end{align}
where\footnote{Both path-integral regularisation and operator ordering in $H$ can generate a purely quantum term proportional to the Ricci scalar $\mathcal{R}(z)$ in the worldline action~\cite{Bastianelli_vanNieuwenhuizen_2006}. $\mathcal{R}=0$ for the Vaidya metric, however, so this subtlety can be ignored. Moreover, the path integral representation is not unique, as pointed out in~\cite{Parker:1979mf} -- our choice corresponds to $p=0$ in the notation of that paper.}
\begin{align}
    S=-\int_{-T/2}^{T/2}\!\ud\tau\, \frac{1}{4}g_{\mu\nu}(z)\dot{z}^{\mu}\dot{z}^{\nu} \;.
\end{align}
Note that, compared to Sec.~\ref{sec:tunnel_EM}, we have made the innocuous shift $\tau\rightarrow \tau-T/2$, which allows to more easily make contact with the geodesics discussed above, in which the proper time runs from $-\infty$ to $\infty$ between asymptotic regions. (Recall from Sec.~\ref{sec:modes_amplitudes} that we only need the asymptotic behaviour of our wavefunctions.)

As in Eq.~(\ref{sol-worldline-def}), we attach the proper-time kernel to some state $\ket{\psi}$, and integrate over all proper time to generate a solution of the wave equation; the worldline representation of mode functions in the Vaidya metric is thus
\begin{align}\label{general-Vaidya-integral}
     \int_{-\infty}^{\infty}\! \ud T \!\int\!\ud^4y\sqrt{-g(y)}\, \psi(y)\int^{z(T/2)=x}_{z(-T/2)=y}\!\mathcal{D} z \, e^{iS} \;.
\end{align}
Factors of the metric determinant appear explicitly both when attaching $\ket{\psi}$ and in the path integral measure. Typically these factors are exponentiated via the introduction of worldline ghosts~\cite{Bastianelli:1991be} but, written in Cartesian form, the Vaidya metric has a constant determinant, so the measure effectively reduces to that in Minkowski and we do not need to invoke ghosts. For a comprehensive discussion see~\cite{Bastianelli_vanNieuwenhuizen_2006}.

Although we can sidestep many technical difficulties in our setup, the event horizon introduces further specific challenges, related to both the definition of the path-integral measure and the choice of `initial' state $\ket{\psi}$. This will become explicit as we proceed to the next step, which is to construct the `out' modes, as used by Hawking~\cite{Hawking:1975vcx}, to find the appropriate Bogoliubov coefficients and thus the number of created particles in the Vaidya spacetime.

%%%%%%%%%%%%%%%%
\subsubsection{The out-modes}\label{subsubsec:out_mode}
Working in the formalism of Sec.~\ref{sec:modes_amplitudes}, the appropriate past asymptotic boundary for massless scalars is $\mathcal{I}^{-}$, from which originate all radial null geodesics. The appropriate future asymptotic boundary is the union of the future event horizon $\mathcal{H}$ and the endpoint $\mathcal{I}^{+}$ of all radially null geodesics that escape the collapse. It is useful to make a separation of the out-modes $\varphi^{\pm\rm out}_\ii$
into two sets, depending on their asymptotic support. We write $\varphi^{\pm\mathcal{I}}_\ii$ for wavefunctions which reduce to positive and negative energy modes near $\mathcal{I}^{+}$, but have no support on $\mathcal{H}$. Similarly, $\varphi^{\pm\mathcal{H}}_\ii$ denotes wavefunctions which reduce to positive and negative energy modes near $\mathcal{H}$, with no support on $\mathcal{I}^+$.
While the $\varphi^{\pm\mathcal{I}}_\ii$ can be uniquely specified, the choice of $\varphi^{\pm\mathcal{H}}_\ii$ involves a degree of arbitrariness. For brevity, we shall now focus on the positive-energy modes, writing $\varphi_\ii^\mathcal{I}\equiv \varphi_\ii^{+\mathcal{I}}$, as the negative-energy ones can be obtained from complex conjugation.

The symmetry of the problem tells us to work in a basis of spherical modes, such that `$\mathbf{i}$' corresponds to the set $(l,m,\epsilon)$, where $l$ and $m$ are the standard angular momentum quantum numbers, while $\epsilon$ is an energy. We will focus on $l=0$ throughout, and so label the modes only by energy, $\ii\to\epsilon$.  Now, by definition, the out modes $\varphi^{\mathcal{I}}_\epsilon$ behave near the asymptotic future null boundary as
\begin{align}
    \varphi^{\mathcal{I}}_{\epsilon}(y)\sim \frac{e^{- i\epsilon U}}{\epsilon R}%{\epsilon r}
    \;, \qquad y\stackrel{\rightarrow}{\in}\mathcal{I}^+ \;.
\end{align}
This suggests that the natural choice of $\ket{\psi}$ is simply
\begin{align}\label{eq:phi_initial_out}
    {\psi(y)}\sim \frac{e^{- i\epsilon U}}{R}
    \;,\qquad \epsilon>0\,,
\end{align}
in which $U$ here is the outgoing null coordinate defined in (\ref{U-def}). Attaching this state will again generate a boundary term in the action, in analogy to the electromagnetic case, we will have the worldline action
\begin{align}\label{eq:def_SWL_hawking}
   S_{\rm WL}=-\int_{-T/2}^{T/2}\frac{1}{4}g_{\mu\nu}(z)\dot{z}^{\mu}\dot{z}^{\nu}d\tau-\epsilon U(-T/2) \;.
\end{align}
The energy-dependent boundary term  can be compared directly to the momentum-dependent boundary term in the corresponding electromagnetic action, see (\ref{eq:worldline-action}).

%%%%%%%%%%%%%%%%%%
\begin{figure}[t!]
    \centering
%%%%%%%%%%%%%%%   
    \begin{subfigure}[t]{0.3\textwidth}
    \includegraphics[width=\textwidth]{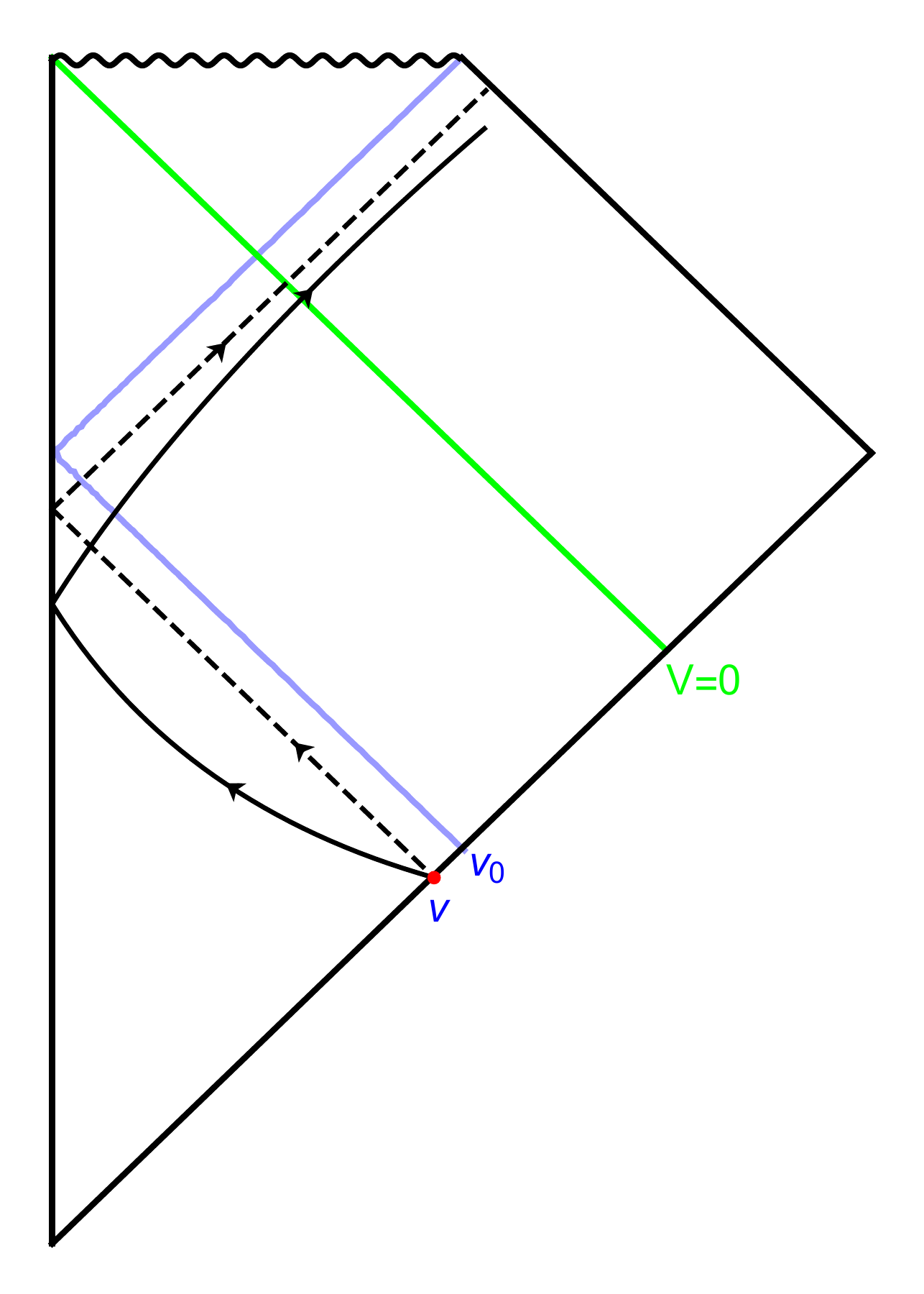}
    \caption{$v<v_0$}
    \label{fig:paths_out_v_lessthan_v0}
    \end{subfigure}
%%%%%%%%%%%%%%
     \hspace{2.5cm}
 %%%%%%%%%%%%%    
    \begin{subfigure}[t]{0.3\textwidth}
    \includegraphics[width=\textwidth]{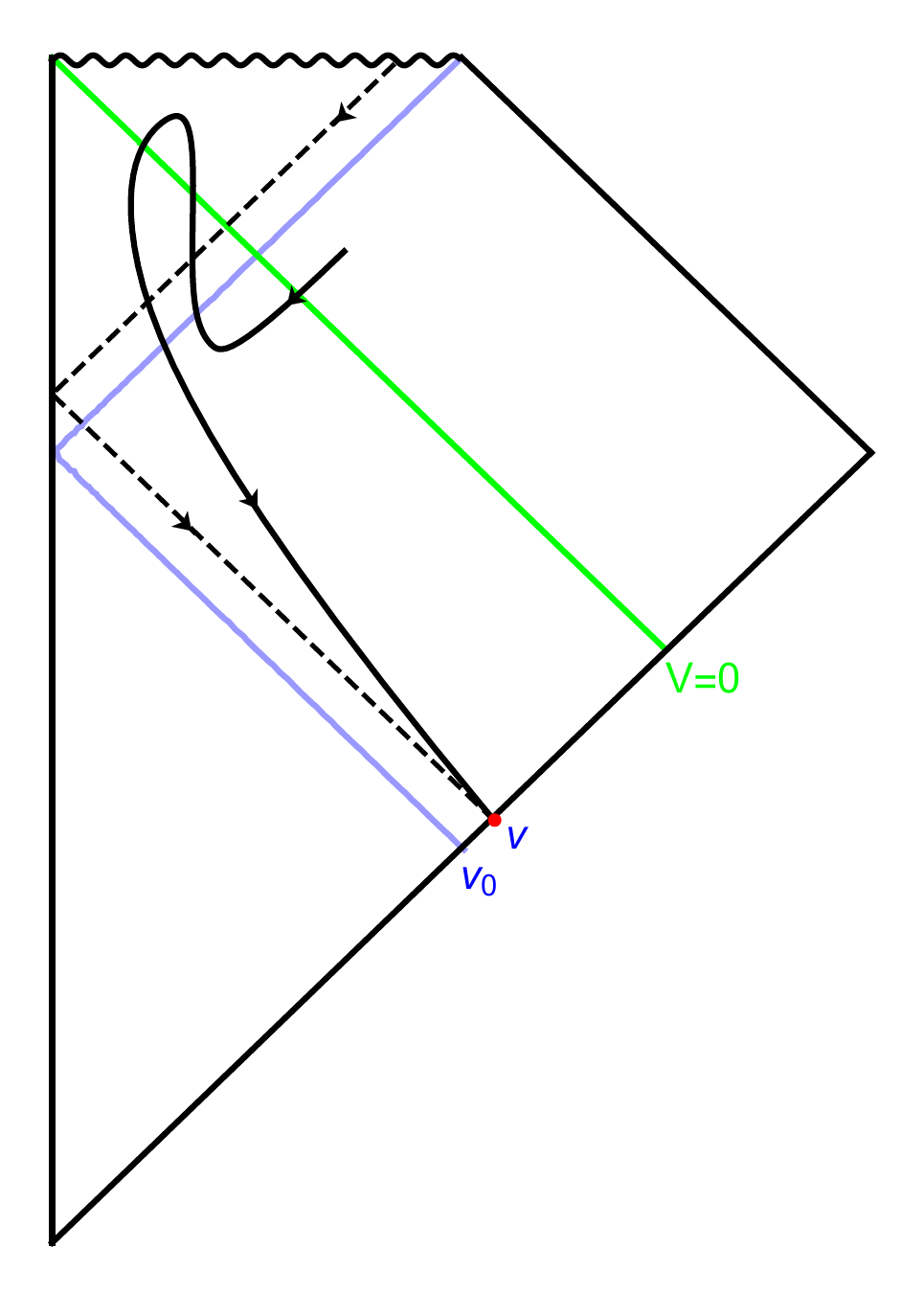}
    \caption{$v>v_0$}
    \label{fig:paths_out_v_greateerthan_v0}
    \end{subfigure}
%%%%%%%%%%%%%    
    \caption{\label{fig:paths_out} The solid black curves illustrate representative worldlines that contribute dominantly to the path integral representation of $\varphi^{\rm F}_{\epsilon}$, while the dashed curve represents the saddle point geodesics.}
\end{figure}
%%%%%%%%%%%%%%%%%%%%%%%%%%%

Now, the out-modes we consider have no support inside the horizon. This suggests that the measure in (\ref{general-Vaidya-integral}) be taken only over worldlines that do not cross the horizon. Indicating this measure by $\mathcal{D}_{+}[z]$, we expect to obtain the out-mode  $\varphi_{\epsilon}^{\mathcal{I}}$ from the path integral
\begin{align}\label{eq:WLPI_for_phi_out}
    \varphi_{\epsilon}^{\mathcal{I}}(x) \overset{?}{=}\int_{-\infty}^{\infty}\!\ud T\int^{z(T/2)=x}\! \mathcal{D}_{+}[z]\, e^{iS_{\rm WL}} \;.
\end{align}
Our aim is to evaluate this path integral, and thus the wavefunction, in the past asymptotic region where the Bogoliubov coefficients are encoded, see~(\ref{eq:outmode_gen}). In doing so, we would in effect be evolving information from the future null boundary back in time to $\mathcal{I}^{-}$, which was Hawking's method of obtaining the Bogoliubov coefficients and particle spectrum.

In Sec.~\ref{sec:tunnel_EM} we were able to evaluate our path-integral exactly, as it was Gaussian. This is not the case here -- evaluating (\ref{eq:WLPI_for_phi_out}) exactly is challenging (part of the broader challenge of evaluating path integrals on bounded domains~\cite{kleinert2006path,chaichian2018path,Vassilevich:2003xt,Bastianelli:2006hq,Bastianelli:2008vh,Asorey:2007zza,Corradini:2019nbb}). Fortunately, this will not prevent us from gaining useful insights;  Hawking argued that the relevant piece of $\varphi^{\rm\mathcal{I}}_{\epsilon}$ can be extracted \emph{within the geometrical optics approximation}. This means that the path integral is well approximated by its semiclassical value, and the saddle points of the integral are of course described by the geodesics studied in Sec.~\ref{subsec:radial_geodesics}.

To investigate the saddle point geodesics more closely, in particular the appropriate boundary conditions,  we vary the action $S_{\rm WL}$ under the assumptions that (i) the initial endpoint $\tau=-T/2$ is in the region $V>0$, where we attach $\ket{\psi}$, and (ii) the final endpoint $\tau=T/2$ is in the region $V<0$, because we are interested in $\varphi_\epsilon^{\mathcal{I}}$ as $x$ approaches $\mathcal{I}^{-}$. The relevant geodesic will then extend from $\mathcal{I}^-$ to $ \mathcal{I}^+$.

Varying the action, we find
\begin{align}\label{variation}
    \delta S_{\rm WL}&=\int_{-T/2}^{T/2}\left[\mathcal{E}_{\mu}\right]\delta z^{\mu}\ud\tau\\ \nonumber
    &+\left.\left[\frac{1}{2}\dot{U}\delta R+\frac{1}{2}\left\{\dot{R}+\left(1-\frac{2GM}{R}\right)\dot{U}-2\epsilon\right\}\delta U\right]\right\rvert_{-T/2}\\\nonumber
    &+\left.\left[\frac{1}{2}\dot{V}\delta R+\frac{1}{2}\left(\dot{R}-\dot{V}\right)\delta V\right]\right\rvert^{T/2}\\\nonumber
    &-\left.\left[\frac{1}{2}R^2\sin^2\theta\, \dot{\phi}\,\delta\phi+\frac{1}{2}R^2\dot{\theta}\,\delta\theta\right]\right\rvert^{T/2}_{-T/2} \,,
\end{align}
where $\mathcal{E}_{\mu}$ denotes the equations of motion. The vanishing of (\ref{variation}) naturally implies Dirichlet conditions on the fluctuation at the `final' point where we evaluate the wavefunction, i.e.~ $\delta z^\mu(T/2)=0$. From here on we will use lower case letters for $V$ and $R$ to refer to the coordinates at the final point, i.e.~$z^\mu(T/2)=x^\mu = (v,r,\phi_0,\theta_0)$, hence the saddle-point worldline of interest obeys
\be
\label{eq:final_conds_Hawking}
    V(T/2) = v\;, \quad
    R(T/2) = r\rightarrow\infty \;, \quad 
    \phi(T/2) = \phi_0 \;, \quad
    \theta(T/2) = \theta_0\;,
\ee
in which the values of $V$ and $R$ capture the limit $x \stackrel{\rightarrow}{\in} \mathcal{I}^- $.  The vanishing of (\ref{variation}) at $\tau=-T/2$, where we attach $\ket{\psi}$, requires us to impose the Neumann outgoing conditions
\be
    \label{eq:initial_conds_Hawking}
   \dot{U}(-T/2) = 0 \;, \quad 
    \dot{R}(-T/2) = 2\epsilon \;,
    \quad
    \dot{\phi}(-T/2)=\dot{\theta}(-T/2)=0\;.
\ee
The conditions on the angular coordinates simply reflect the fact that the wavefunction corresponds to an \( l = 0 \) spherical mode. {(Note that because we attached $\psi$ at $-T/2$ rather than $+T/2$,
the terms `initial' and `final' are used in a formal sense.)} 

Thus, in the semiclassical limit, the result of performing our path integral will be equal to the exponent of the classical action, evaluated on a worldline $z_{\rm cl}(\tau)$ respecting the boundary conditions \eqref{eq:initial_conds_Hawking} and \eqref{eq:final_conds_Hawking}. We then have to perform the $T$-integral, which in the saddle-point approximation will select out the value of $T$ such that $z_{\rm cl}$ becomes a classical geodesic. 

However, since the geodesic is null, $\dot{z}^{\mu}_{\rm cl}\dot{z}_{\mu\, \rm cl}=0$, the bulk term of (\ref{eq:def_SWL_hawking}) vanishes so that the classical action receives a contribution only from the boundary term $\epsilon U(-T/2)$ which, by (\ref{eq:final_conds_Hawking}), will be $T$-independent. There will be a $T$-dependent fluctuation determinant coming from expanding around saddle-points, but we can neglect this;  the out-mode, in the geometrical optics limit, will take the form
\begin{align}
    \varphi_{\epsilon}^{\mathcal{I}}(x)\propto\frac{e^{iS_{\rm WL}[z_{\rm cl.}]}}{\epsilon r} \;, \qquad x\stackrel{\rightarrow}{\in}\mathcal{I}^- \;,
\end{align}
in which the factor of $1/r$ will arise, as in standard quantum mechanics, from the integration of the `initial' angular coordinates that give rise to $l=0$ spherical modes, see~\cite{kleinert2006path} -- we have included this as a reminder that the wavefunction is being evaluated asymptotically.

We have already calculated the asymptotic value of $U$ for an initially ingoing radial geodesic in Sec.~\ref{subsec:radial_geodesics}. Using the result therein, we have that the out-mode, evaluated near $\mathcal{I}^-$, has the form
\begin{align}\label{eq:f_near_I_minus}
    \varphi^{+\mathcal{I}}_{\epsilon}(x)\sim \frac{e^{-i\epsilon\left[v-4GM\log\left(
    \frac{v-v_0}{v_0}
    \right)
    \right]}}{\epsilon r}\Theta(v_0-v)
    \;, \qquad x\stackrel{\rightarrow}{\in}\mathcal{I}^- \;,
\end{align}
The Heaviside theta function (c.f.~(\ref{classical-like-solution}) in the Schwinger effect) here reflects that for \( v > v_0 \), there are no saddle points (geodesics) that both satisfy the boundary conditions and remain outside the horizon.  These are the out-modes as used by Hawking~\cite{Hawking:1975vcx}, which we have recovered using the worldline approach. The fact that our result only depends on the boundary term serves as a worldline-based proof of Hawking’s intuitive argument that the asymptotic form of the out-mode at $\mathcal{I}^{-}$ can be derived from the asymptotic value of $U$ at $\mathcal{I}^+$ associated with a radial geodesic that initially starts at $V = v$ on $\mathcal{I}^{-}$. 

Finally, the spherical in-modes, on the other hand, take the form
\begin{align}
    \varphi^{\pm\rm in}_{\epsilon}(x)\sim \frac{e^{\mp i\epsilon v}}{\epsilon r}
    \;,
    \qquad
    x\stackrel{\rightarrow}{\in}\mathcal{I}^-\,,
\end{align}
and so the Bogoliubov coefficients can be found simply by Fourier transforming~ \eqref{eq:f_near_I_minus}. The particle spectrum, recall by~\eqref{eq:particle_num_gen}, then takes the well-known form of a thermal distribution
\begin{align}
    \braket{{\rm in}|a_{{\mathcal{I}}}^{\epsilon\,\dagger}a_{\mathcal{I}}^{\epsilon}|{\rm in}}\propto\frac{1}{e^{8\pi GM\epsilon}-1} \;.
\end{align}

\subsection{The tunnelling modes and Hawking radiation}
We finally turn the the tunnelling modes, from which we will read off the amplitude for Hawking radiation. The tunnelling modes have support on $\mathcal{H}$ as well as $\mathcal{I}^{+}$. Their worldline representation should then allow for paths that cross the horizon, c.f.~the discussion below (\ref{full-tunnelling-phi}). This implies that  we also need a state $\psi$ which can be defined beyond the horizon. If we want to attach the `same' state as for the out modes, (\ref{eq:phi_initial_out}), then we need only to find a way to continue the coordinate $U$. 

The way to do so is provided by Sec.~\ref{sec:modes_amplitudes} and Sec.~\ref{sec:tunnel_EM}. The first line of (\ref{eq:def_Feynman_modes}) tells us that we need to continue $U$ such that the resulting tunnelling mode is a positive energy function on $\mathcal{I}^-$. The intuition for \emph{how} to do this is provided by the Schwinger-effect examples in (\ref{Fourier1}) and (\ref{Fourier2}) -- we define $U$ for the entire range of $0<R<\infty$ by the same analytic continuation as used for the Schwinger effect, i.e.~we have (\ref{U-def}) but with log understood to be take its principal value.
Our initial state may then again be written
\begin{align}\label{eq:phi_initial_out}
    \psi(y)
    \propto \frac{e^{-i\epsilon U}}{\epsilon R} \;, \qquad \epsilon>0\,.
\end{align}
It follows that the appropriate worldline action $S_{\rm WL}$ also has the same form as for the out-modes, see \eqref{eq:def_SWL_hawking}, and we can write
\begin{align}\label{eq:WLPI_for_phi_F}
    \varphi_{\epsilon}^{\text{F}}(x)=\int_{-\infty}^{\infty}dT\int^{z(T/2)=x} \mathcal{D}[z] e^{iS_{\rm WL}} \;,
\end{align}
where the measure now allows for worldlines that may cross the horizon. The boundary conditions on the worldlines also remain the same as in~\eqref{eq:initial_conds_Hawking} and \eqref{eq:final_conds_Hawking}. This is because the initial condition~\eqref{eq:initial_conds_Hawking} specifies only the proper-time derivative $\dot{U}(-T/2)$ and not $U(-T/2)$ itself. It follows that the geometrical optics approximation to the (positive and negative energy) tunnelling modes will, as for the out-mode, take the form
\begin{align}\label{eq:WLPI_semiclass_F}
    \varphi_{\epsilon}^{\pm\text{F}}(x)\propto\frac{e^{\pm iS_{\rm WL}[z_{\rm cl.}]}}{\epsilon r} \;,
\end{align}
where $z_{\rm cl}(\tau)$ is again the classical geodesic respecting the boundary conditions~\eqref{eq:final_conds_Hawking} and \eqref{eq:initial_conds_Hawking}.  This means that for $v<v_0$ the tunnelling and out-modes have the same functional form (just as the (\ref{full-tunnelling-phi}) and in-modes (\ref{classical-like-solution}) for the Schwinger effect agree outside the horizon). So, we have,  from now on focussing on the \emph{negative} energy mode which, see (\ref{eq:def_Feynman_modes}), most directly encode the pair production amplitude, 
\begin{align}\label{eq:phi_F_classical_region}
    \varphi_{\epsilon}^{-\text{F}}(x)\propto\frac{e^{+i\epsilon\left[v-4GM\log\left(\frac{v-v_0}{v_0}\right)\right]}}{\epsilon r}= \varphi^{-\mathcal{I}}_{\epsilon}(x)\;, \qquad x\stackrel{\rightarrow}{\in}\mathcal{I}^- \quad\text{and}\quad v<v_0.
\end{align} 
For $v>v_0$ on the other hand, while no saddle points contribute to $\varphi^{\mathcal{I}}_{\epsilon}(x)$, the tunnelling mode receives contributions from saddle points (geodesics) that cross the horizon. The final result for the tunnelling wavefunction near $\mathcal{I}^-$ is again determined by the value of $U(-T/2)$, but this is now complex. One finds
\begin{align}\label{eq:phi_F_tunl_region}
 \varphi_{\epsilon}^{-\text{F}}(x)
 \propto
 e^{-4\pi GM\epsilon}
 \frac{e^{+i\epsilon\left[v-4GM\log\left(\frac{v_0-v}{v_0}\right)\right]}}{\epsilon r} \;, \qquad x\stackrel{\rightarrow}{\in}\mathcal{I}^- \;,\qquad v >v_0 \;, 
\end{align}
in which the leading exponential factor arises from the imaginary part of $U(-T/2)$. To interpret this result, we recall the discussion in Sec.~\ref{subsec:radial_geodesics}, from which one might expect the contributing saddle points to here describe geodesics that start from $\mathcal{I}^{-}$, cross the horizon, and ultimately hit the singularity. However, \eqref{eq:constraints_geodesic} tells us that for such a geodesic, with $v>v_0$, the outgoing energy would be negative, violating the boundary condition $\dot{R}(-T/2)=2\epsilon>0$ in \eqref{eq:initial_conds_Hawking}. The correct interpretation of the contributing saddle points is rather that they are the \emph{time-reversed} versions of the horizon-crossing radial geodesics -- they start from the singularity, cross the horizon, and ultimately reach $V=v$ at $\mathcal{I}^{-}$, see Fig.~\ref{fig:paths_out_v_greateerthan_v0}.  These time-reversed geodesics satisfy all the boundary conditions in \eqref{eq:final_conds_Hawking} and \eqref{eq:initial_conds_Hawking}, hence are valid saddle points of the path integral in \eqref{eq:WLPI_semiclass_F}.

For this reason we identify (\ref{eq:phi_F_tunl_region}), aside from the leading real exponential, with the asymptotic form of the wavefunction $\varphi^{-\mathcal{H}}_{\epsilon}(x)$; this is the mode function having support only on $\mathcal{H}$ in the future, and the minus sign appears in the superscript because the saddle point is a time-reversed geodesic, hence this function should be associated to a negative energy state.

We can summarise the situation as follows. For $x$ approaching $\mathcal{I}^-$ the tunnelling wavefunction is a negative energy function with the form
\begin{align}\label{eq:varphi_F_Iminus}
    \varphi_{\epsilon}^{-\rm F}(x) \sim \frac{1}{\epsilon r}
    \begin{cases}
    \displaystyle {e^{+i\epsilon\left[v-4GM\log
    \left(\frac{v-v_0}{v_0}\right)\right]}}&  v < v_0 \\[10pt]
    \displaystyle
    e^{-4\pi GM\epsilon} {e^{+i\epsilon\left[v-4GM\log\left(\frac{v_0-v}{v_0}\right)\right]}} & v> v_0
    \end{cases} \qquad x\stackrel{\rightarrow}{\in}\mathcal{I}^- \;.
\end{align}
Expressed as a function \emph{throughout} spacetime, the tunnelling wavefunctions will take the form
\begin{align}\label{eq:varphi_F_full}
   \varphi_{\epsilon}^{\rm\pm F}(x)=\varphi_{\epsilon}^{\pm\mathcal{I}}(x)+e^{-4\pi GM\epsilon}\varphi^{\mp\mathcal{H}}_{\epsilon}(x) \;,
\end{align}
in which $\varphi^{\pm\mathcal{I}}_\epsilon$ are postive/negative energy functions on $\mathcal{I}^+$ while $\varphi_\epsilon^{\mp\mathcal{H}}$ are negative/positive energy function on the horizon. We have not evaluated these functions for all arguments, nor should be need to, but we should verify that our $\varphi_{\epsilon}^{-\rm F}(x)$ is indeed a valid Feynman mode. This we do by checking against the key properties discussed at the end of Sec.~\ref{sec:modes_amplitudes} below (\ref{eq:def_Feynman_modes}).

First, in the past, $\varphi^{-\text{F}}_{\epsilon}(x)$ takes the form given in \eqref{eq:varphi_F_Iminus}, the $v$-dependence of which closely resembles the $x^\LCm$-dependence of the tunnelling wavefunction $\varphi_{p}(x)$ in the Schwinger case. As for that example, see \eqref{Fourier2}), we find that $\varphi^{\pm\text{F}}_{\epsilon}(x)$ is a positive/negative energy mode as $x$ approaches $\mathcal{I}^{-}$. Next, to analyse the behaviour of $\varphi^{\pm\text{F}}_{\epsilon}(x)$ near $\mathcal{I}^{+}\cup \mathcal{H}$, observe that the global form given in \eqref{eq:varphi_F_full} makes it clear that the tunnelling wavefunction asymptotes to the free spherical out-mode near $\mathcal{I}^{+}$. Near the horizon, however, it asymptotes to a negative-energy horizon mode. This establishes that $\varphi^{\pm\text{F}}_{\epsilon}(x)$ is indeed a Feynman mode.

We can now identify the amplitude for pair creation, which we identify with
\begin{align}
    \mathcal{A}_{0\rightarrow 2}(\epsilon,\epsilon')\equiv \frac{\braket{{ \rm in}|a_{\mathcal{H}}^{\epsilon'}\,a_{\mathcal{I}}^{\epsilon}\mathcal{S}|{\rm in}}}{\braket{{ \rm in}|\mathcal{S}|{\rm in}}} 
\end{align}
where $a_{\mathcal{H}}^{\epsilon}$ and $a_{\mathcal{I}}^{\epsilon}$ are annihilation operators associated to the modes $\varphi_{\epsilon}^{\mathcal{I}}$ and $\varphi_{\epsilon}^{\mathcal{H}}$ respectively. The amplitude can now be read off from (\ref{eq:def_Feynman_modes}) and (\ref{eq:varphi_F_full}) as, up to a phase,
\begin{align}\label{eq:amplitude_Hawking}
    \mathcal{A}_{0\rightarrow 2}(\epsilon,\epsilon') =  2\pi {\delta}(\epsilon-\epsilon')\, e^{-4\pi GM\epsilon}\;,
\end{align}
which exhibits both the expected conservation of energy and exponential dependence on the gravitational coupling.

Having drawn parallels with the Schwinger effect throughout this derivation of the amplitude for Hawking radiation, we finish  by highlighting one notable difference between the gauge and gravitational results.
The difference lies in the nature of the worldlines associated with the tunneling processes.

Recall that in the Schwinger effect, the wavefunction in the classically forbidden region $x^\LCm > x^\LCm_h$ is effectively described by a \textit{complex} instanton (which takes the form of two real trajectories connected by a complex part). The analogous contribution to the Feynman wavefunction for Hawking radiation at $v>v_0$ is, in contrast, described by a \textit{real} radial null geodesic propagating backward in time.

In both scenarios the semiclassical exponential suppression associated with tunneling arises from the imaginary part of the classical action, although the details again differ. In the Schwinger case, the imaginary contribution stems from the term $(p_{\perp}^2+m^2)T$  in the worldline action. In the Hawking case, on the other hand, the imaginary part originates from the boundary term $\epsilon U$ which becomes imaginary even though the saddle-point worldline itself remains real. Had we formulated the path integral in a coordinate system adapted to observers outside the horizon, using, say, \emph{outgoing} Eddington-Finkelstein coordinates, the saddle-point radial null geodesic contributing to the tunneling process would naturally arise as a complex solution to the geodesic equation, since it would extend beyond the domain of the coordinate chart. A more covariant perspective is to observe that the boundary term possesses a branch cut across the horizon, hence why real worldlines crossing the horizon can yield imaginary parts. (In the Schwinger case the branch structure only appears after evaluating the $T$-integral.)

\section{Conclusions}\label{sec:conclusion}
%%%
Classically, the dynamics of a charged particle in a constant electric field and those of a massless particle in a black hole spacetime are both influenced by the presence of an (in the former case effective) horizon. Quantum mechanically, there is tunneling at this horizon, reflecting the underlying process of particle creation.

We have applied the Lorentzian worldline path integral approach to describe particle creation via the Schwinger effect, and Hawking radiation 
at the semiclassical level. Our approach was based on a worldline construction of solutions of the appropriate wave equation for a particle in an electric field or in a Vaidya (collapse) spacetime.

The Schwinger effect is nearly fully analytically tractable,
but a general analysis of the solutions of background-coupled wave equations shows that scattering amplitudes, in both gravity and gauge theory, are encoded in their asymptotic behaviour. (Specifically, the asymptotic forms of certain solutions correspond to on-shell Fourier transforms of the amplitudes.) Having only to evaluate our wavefunctions in asymptotic regions, and being guided by results from the Schwinger effect, allowed us to extract the pair-creation amplitude associated with Hawking radiation from 
a Lorentzian worldline path integral.

In both cases, the boundary conditions on the worldlines contributing to the wavefunction admit a simple interpretation: one end satisfies a Dirichlet condition, aligning the worldline coordinate with the argument of the wavefunction, while the other end imposes a condition on the worldline’s energy-momentum, matching that of the wavefunction. The latter condition can be implemented by supplementing the worldline action with an appropriate boundary term. At the semiclassical level, our calculations reduced to the evaluation of the worldline action on classical trajectories satisfying these boundary conditions.

When the wavefunction is evaluated in classically forbidden regions, the worldline action acquires an imaginary part, giving rise to an exponentially suppressed factor akin to tunneling in quantum mechanics.  Interestingly, while the relevant worldline trajectories are complex in the Schwinger effect, they remain real in the case of Hawking radiation, despite the action acquiring an imaginary part.

The horizon present already in the classical theory manifests as a branch cut in the wavefunction. We showed that the worldline formalism naturally bridges these classical and quantum features in both the Schwinger and Hawking cases. The wavefunctions in these two cases exhibit similar non-analytic behaviour (near critical values of their arguments associated with the horizons), but this arises in distinct ways in our analysis. In the Schwinger case, the branch cut emerges from the proper-time integral, whereas in the Hawking case, it can be traced back to the branch cut structure of the boundary term in the worldline action.

Our analysis opens up several directions for further research.
Our treatment of Hawking radiation relied on a semiclassical approximation; it would be interesting to see if our results could be derived from a more rigorous formulation of the worldline path integral in bounded spacetimes, or how exact solutions to the wave equation in Schwarzschild spacetime, discussed in \cite{Li:2016sjq}, emerge from the worldline formulation.
For the Schwinger effect, we showed that the complex worldline instanton responsible for tunneling effectively represents a family of spacelike worldlines contributing to the wavefunction.
An interesting question is whether a similar structure exists in the Hawking case — is there a set of real worldlines with real action can capture the tunneling behavior of the Feynman mode?
One avenue worth exploring is the potential role of non-radial geodesics in the computation of fixed-angular-momentum modes, and whether they provide the appropriate description.
Finally, another possible extension of our results is to cosmological pair creation, with de Sitter spacetime serving as an obvious starting point, due to the presence of a cosmological horizon. For very recent work on this topic see~\cite{GregerNew}.

\medskip

\textit{We thank Rafa Aoude, Gerald Dunne, Holger Gies and Christian Schubert for many useful discussions. We thank Philip Semr\'en and Greger Torgrimsson both for useful discussions and for sharing a draft of~\cite{GregerNew}. The authors are supported by the STFC Consolidated Grant ``Particle Theory at the Higgs Centre" ST/X000494/1 (AI) and the EPSRC Standard Grant EP/X024199/1 (AI, KR).}

\appendix

\section{Asymptotics and direct worldline computation of amplitudes in constant electric fields}\label{app:Schwinger}
One purpose of this appendix is to show that particle scattering and pair production amplitudes in a constant background are supported on spacetime region outside, respectively beyond, a classical horizon. In addition, however, this section also presents a direct computation of scattering amplitudes in a constant external background, employing the worldline path integral approach. In particular, we demonstrate how these amplitudes can be derived without any recourse to properties of complicated parabolic cylinder functions that commonly appear in treatments of this system.

In terms of in/out modes and the propagator $G(x,y)$, the pair creation amplitude has the LSZ expression~\cite{Barut:1989mc}
\begin{align}\label{app-amp-02}
    \mathcal{A}_{0\rightarrow 2}(q,p)&= \lim_{t_b=t_a\rightarrow\infty}
    \int\!\ud^3\mathbf{y}\ud^3\mathbf{x}
    \,\phi^{\dagger \rm out}_{q}(t_a,\mathbf{x})
    {\overset{\leftrightarrow}{\partial}}_{t_a}
    G(x,y){\overset{\leftrightarrow}{\partial}}_{t_b}
    \phi^{\dagger \rm out}_{p}(t_b,\mathbf{y})\,,
\end{align}
while the 1$\rightarrow$1 amplitude is
\begin{align}\label{app-amp-11}
    \mathcal{A}_{1\rightarrow 1}(p\rightarrow q)&=\lim_{t_b=-t_a\rightarrow\infty}
    \int\!\ud^3\mathbf{y}\ud^3\mathbf{x}
    \,\phi^{\dagger \rm out}_{q}(t_a,\mathbf{x})
    {\overset{\leftrightarrow}{\partial}}_{t_a}
    G(x,y){\overset{\leftrightarrow}{\partial}}_{t_b}
    \phi^{\rm in}_{p}(t_b,\mathbf{y})\,,
\end{align}
both being naturally expressed in terms of `instant' time $x^0=t$. We therefore switch here to the commonly-used time-dependent gauge $eA_{\mu}=(0,0,0,eE t)$ (in Cartesian coordinates). Note that we only need the in/out modes in the large-time limit; these asymptotic solutions to the Klein-Gordon equation are easily found to be
\begin{align}
    \phi^{\rm in}_{p}(x)&\sim \frac{e^{ieEt^2}}{\sqrt{-eEt}}e^{-ip_jx^j}\qquad;\qquad t\rightarrow-\infty\\
    \phi^{\rm out }_{p}(x)&\sim \frac{e^{-ieEt^2}}{\sqrt{eEt}}e^{-ip_jx^j}\qquad;\qquad t\rightarrow\infty
\end{align}
As a result, the amplitudes (\ref{app-amp-02}) and (\ref{app-amp-11}) differ only in the sign of $p_\mu$ in the wavefuncion at $\mathbf y$, and the sign of $t_a$. This allows us to perform the bulk of the calculation for both cases simultaneously. Writing $\mathcal{A}_{0\to2}\equiv \mathcal{A}_\LCp$ and $\mathcal{A}_{1\to1}\equiv \mathcal{A}_\LCm$ and using the path-integral representation of the propagator, we have
\begin{align}\label{eq:amp_pm_PI}
    \mathcal{A}_\pm &=\lim_{t_a\rightarrow\infty}\int\!\ud^3\mathbf{x}\int_{0}^{\infty}\!\ud T\,
    \mathcal{F}_\pm(T)\int^{z(T)=(t_a,\mathbf{x})}_{z^0(0)=\pm t_{a}} \mathcal{D}[z]\, e^{iS_\pm}\,,
\end{align}
in which the relevant action is
\begin{align}
    S_\pm =-\int_{0}^{T}\left(\frac{1}{4}\dot{z}^2+eA_{td}\cdot \dot{z}+m^2\right)\ud\tau +q_i x^{i} \pm p_{i}z^{i}(0)\,,
\end{align}
while the prefactor takes the form
\begin{align}
    \mathcal{F}_\pm(T)=e^{ieE\bar{t}^2_a}\times\frac{1}{2}eE\bar{t}_a[1+\tanh^{\pm 1}(eET)]^2\,.
\end{align}
with $\bar{t}_{a/b}=t_{a/b}-p_3/eE$. The worldline path integral is Gaussian and hence can be performed exactly, yielding
\begin{align}
    \mathcal{A}_\pm 
    &=
    \lim_{t_a\rightarrow\infty}\int\!\ud^3\mathbf{x}
    \int_{0}^{\infty}\!\ud T\,
    \left(\frac{eE}{2\pi i\sinh(2eE T)}\right)^{1/2}\mathcal{F}_\pm(T)\, e^{iS_\pm (z_\text{cl})}\;,
\end{align}
in which  \begin{align}
   S_\pm(z_\text{cl}) =-(p_{\perp}^2+m^2)T + 
    (q\pm p)_{i}x^{i}-eE\,\bar{t}_a^2\tanh^{\pm 1}(eE T)\;,
\end{align}
is indeed the classical action evaluated on the saddle point worldline satisyfing the boundary conditions
\begin{align}
    z^{\mu}(T)&=(t_a,\mathbf{x})\,,\\
    z^{0}(0)&= \pm t_a\quad;\quad 
    \frac{1}{2}\dot{z}^{3}(0)-eE z^{0}(0)= \pm p^{3}\quad;\quad
    \frac{1}{2}\dot{z}^{\perp}(0)= \pm p^{\perp}\,.
\end{align}
From here it is notationally simpler to focus on the amplitudes separately.

\subsection{The 0-to-2 amplitude}
The explicit form of the saddle point worldline in the pair production amplitude is the same as that found for the computation of the tunnelling wavefunction in the text, up to some transformation between integration constants. Specifically, and reverting to lightfront coordinates, we have here
\begin{align}\label{class-sol-for-pairs}
    {z}_{\rm cl}^{-}(\tau)&= \hat{x}_h^- +\left({x}^{-}-\hat{x}^-_h\right)e^{2eE(T-\tau)} \;, \qquad
     {z}_{\rm cl}^{+}(\tau)={x}^{+}+\left(\frac{e^{2eE\tau}-e^{2eET}}{eE}\right)\hat{p}_{-}\,,
\end{align}
in which, note, ${x}^{\pm} =\frac{1}{\sqrt{2}}(t_a\pm x^{3})$, while we define $\hat{x}^\LCm_h$ and $\hat{p}_\LCm$ by
\begin{align}\label{eq:xm-xh}
    {x}^- -\hat{x}_h^- = \frac{\sqrt{2}}{eE} \frac{(e Et_a-p_3)}{1+e^{2eE T}}=\frac{\hat{p}_{-}}{eE}\,.
\end{align}
Since $t_a\rightarrow \infty$, in the computation of $\mathcal{A}_{0\rightarrow 2}$, we clearly have ${x}^--\hat{x}_h^->0$, which is analogous to the tunnelling region in our analysis of $\varphi_{p}(x)$ in the text. Morevover, these worldlines are spacelike, $\dot{z}_{\text{cl}}^2 <0$, as can be easily verified (compare with the worldlines contributing to $\varphi_{p}(x)$ when $x^{\LCm}>x^{\LCm}_{h}$).

To compute the $T$-integral, we can take advantage of the ${t}_{a}\rightarrow \infty$ limit. To this end, we first define the new variable
\begin{align}
    e^{-eET}=\left(\frac{1}{\bar{t}_a}\sqrt{\frac{\lambda}{2eE}}\right)s\,,
\end{align}
where $\lambda=(p_{\perp}^2+m^2)/eE$, so that the amplitude becomes
\begin{align}
    \mathcal{A}_{0\rightarrow 2}(q,p)&=\lim_{t_a\rightarrow\infty}\hat{\delta}(\mathbf{p}+\mathbf{q})\left(\frac{\lambda}{2eE\bar{t}_a^2}\right)^{\frac{i\lambda}{2}}\left(\frac{2(-1)^{3/4}\sqrt{\lambda}}{\sqrt{\pi}}\right)\,\,\int_{0}^{\infty}\ud s\, e^{i\lambda s^2}\,s^{i\lambda}\,.
\end{align}
The final integral can be performed in terms of the Gamma function, and we arrive at
\begin{align}
    \mathcal{A}_{0\rightarrow 2}(q,p)&=\lim_{t_a\rightarrow\infty}\hat{\delta}(\mathbf{q}+\mathbf{p})\left[-\frac{(2eE\bar{t}_a^2)^{-i\frac{\lambda}{2}}}{\sqrt{2\pi}}\Gamma\left(\frac{1}{2}+\frac{i\lambda}{2}\right)\right]\,.
\end{align}
This matches with the expression derived in~\cite{Barut:1989mc}. In particular, stripping off the delta functions and writing $\mathcal{A}_{0\rightarrow 2}(q,p)=\hat{\delta}(\mathbf{q}+\mathbf{p})\mathcal{M}_{0\rightarrow 2}(q,p)$ it is easily verified that
\begin{align}\label{mod-m-sq-pairs}
    |\mathcal{M}_{0\rightarrow 2}(q,p)|=\frac{e^{-\pi\lambda}}{1+e^{-\pi\lambda}} \;.
\end{align}
Before moving on to the 1-to-1 amplitude we note that in terms of the variable $s$, \eqref{eq:xm-xh} becomes
\begin{align}
    {x}^--\hat{x}_h^-=\frac{s^2\lambda}{\sqrt{2}eE\bar{t}_a}+\mathcal{O}(\bar{t}_a^{-3})>0\,.
\end{align}
which will be useful below. Note that, in contrast to many approaches to the calculation of (\ref{mod-m-sq-pairs}), we have at no point had to invoke any use of parabolic cylinder functions (the exact solutions of the Klein-Gordon equation in a constant electric field). This is due to (i) using from the start that the amplitude is an asymptotic quantity and (ii) using a worldline expression for the propagator. 

\subsection{The 1-to-1 amplitude}
The 1-to-1 amplitude can be obtained from the pair amplitude by making the replacements $eET\rightarrow eET+i\pi/2$ and $p_i\rightarrow -p_i$. Using the same change of variable as above, the one to one amplitude becomes
\begin{align}
    \mathcal{A}_{1\rightarrow 1}(p\rightarrow q)&=\lim_{t_a\rightarrow\infty}\hat{\delta}(\mathbf{q}-\mathbf{p})\left(\frac{\lambda}{2eE\bar{t}_a^2}\right)^{\frac{i\lambda}{2}}\left(\frac{2(-1)^{3/4}\sqrt{\lambda}}{\sqrt{\pi}}\right)\,\,\int_{0}^{\infty}e^{-i\lambda s^2}\,s^{i\lambda}\,ds\,.
\end{align}
We can rotate $s\rightarrow se^{-i\pi/2}$ to convert the remaining integral to that  encountered in the pair-creation amplitude. Hence we immediately arrive at
\begin{align}
    \frac{\mathcal{A}_{0\rightarrow2}(-q,p)}{\mathcal{A}_{1\rightarrow 1}(p\rightarrow q)}=e^{-\frac{\pi\lambda}{2}}e^{i\frac{\pi}{2}} \;,
\end{align}
and hence the nontrivial part of the amplitude obeys
\begin{align}
    |\mathcal{M}_{1\rightarrow 1}(p\rightarrow q)|^2=\frac{1}{1+e^{-\pi\lambda}}\,.
\end{align}
Under the rotation $s\rightarrow se^{-i\pi/2}$ (or equivalently $eET\rightarrow eET+i\pi/2$ and $p_i\rightarrow -p_i$), we also obtain the classical solution ${z}_{\rm cl}(\tau)$ relevant to the 1-to-1 amplitude from that in (\ref{class-sol-for-pairs}). In particular, we find that the analogue of $({x}^--\hat{x}_h^-)$, i.e.~the coefficient of $e^{2eE(T-\tau)}$ in ${z}^\LCm_{\rm cl}(\tau)$, satisfies
\begin{align}
    {x}^--\hat{x}_h^- \to -\frac{s^2\lambda}{\sqrt{2}eE\bar{t}_a}+\mathcal{O}(\bar{t}_a^{-3})<0\,,
\end{align}
showing clearly that the worldlines for the one-to-one amplitude are analogous to those lying outside the horizon in our discussion of $\varphi_{p}(x)$.

%%%%%%%%%%%%%%%%%%%%%
\section{Direct worldline computation of the semiclassical Hawking radiation amplitude}
%%%%%%%%%%%%%%%%%%%%
%
Paralleling the discussion in the previous appendix, we outline here a direct computation of the Hawking pair creation amplitude in the worldline formalism. This will make the connection with the Schwinger effect even more transparent.

The natural analogue of \eqref{eq:amp_pm_PI}, i.e.~the amplitude for Hawking radiation, is
\begin{align}\label{eq:Hawking_ampl_WLPI}
    \mathcal{A}_{0\rightarrow 2}(\epsilon,\epsilon') &=\int_{0}^{\infty} dT \mathcal{F}(T)\int_{\mathcal{E}(\tau_i)=\epsilon'}^{\mathcal{E}(\tau_f)=\epsilon} \mathcal{D}[z]\exp\left[i S\right]\,,
\end{align}
where `$\mathcal{E}(\tau_f)=\epsilon$' is shorthand for the conditions
\be
    \dot{R}(\tau_f)=2\epsilon \qquad\text{and}\qquad \dot{U}(\tau_f)=0 \;,
\ee
(analogously $\mathcal{E}(\tau_i)=\epsilon'$)  which fix the energy, while $\{\tau_{i},\tau_{f}\}$ are appropriate asymptotic values of the affine parameter $\tau$. In the semiclassical limit, where we take both the worldline and proper-time integrals in their saddle-point approximations, the amplitude becomes
\begin{align}\label{eq:Hawking_ampl_WLPI_semicl}
  \mathcal{A}_{0\rightarrow 2}(\epsilon,\epsilon')\sim\exp\left[i \bar{S}_{\rm WL }\right]\Bigg\rvert_{z\rightarrow \bar{z}_{cl}}  \;,
\end{align}
where $\bar{z}_{cl}$ is the radial geodesic satisfying the boundary conditions \eqref{eq:Hawking_ampl_WLPI}, which in turn are implemented by adding appropriate boundary terms to the action as in the text, thus
\begin{align}\label{eq:def_SWL_hawking_2}
   \bar{S}_{\rm WL}=-\int_{\tau_i}^{\tau_f}\frac{1}{4}g_{\mu\nu}(z)\dot{z}^{\mu}\dot{z}^{\nu}d\tau+\epsilon U(\tau_f)-\epsilon'U(\tau_i) \;.
\end{align}
Just as in our analysis of the wavefunction $\varphi^{\mathcal{I}}(x)$ in Sec.~\ref{subsubsec:out_mode}, the action receives contributions only from these boundary terms, which depend only on the asymptotic values of $U$.

\begin{figure}[t!]
    \centering
    \includegraphics[width=.4\textwidth]{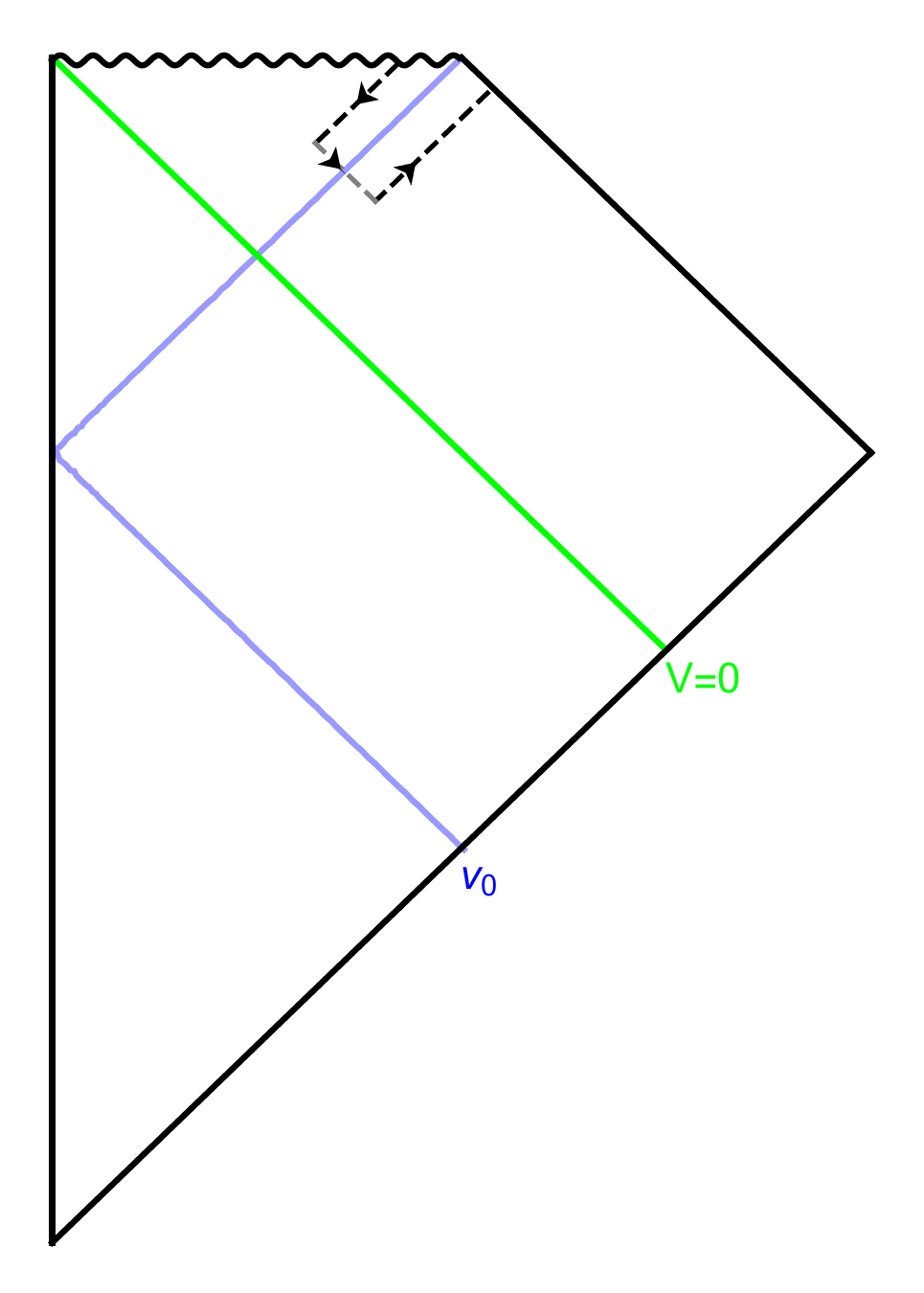}
    \caption{A representative worldline instanton describing Hawking radiation. The thick-dashed lines are real `outgoing' geodesics, while the gray-dashed `ingoing' line represents the real part of the complex wordline instanton.}
    \label{fig:Hawking_instanton}
\end{figure}

In order to motivate the worldline instanton corresponding to the above boundary condition, we first rewrite the asymptotic form of the geodesic solution in \eqref{V-U-geodesic} as follows:
\begin{align}\label{V-U-geodesic_2}
V(\lambda)=4 GM \left(e^{\frac{\lambda }{4 GM}}+1\right)+\lambda +u\quad;\quad    R(\lambda)=2 GM\left( e^{\frac{\lambda }{4 GM}}+1\right) \;,
\end{align}
in which we have defined a non-affine parameter $\lambda$ by $\tau\equiv\frac{GM}{\epsilon}e^{\frac{\lambda(\tau) }{4 GM}}$. Note the striking similarity between $R(\lambda)$ and the instanton $z^\LCm_{\rm cl}(\tau)$ appearing in the Schwinger effect, \eqref{eq:z_classical_expl.}. This suggests that, in analogy to the Schwiner effect, the worldline instanton relevant here is obtained by complexifying $\lambda$ such that it runs from $\lambda(\tau_i)=-i4GM\pi$ to $\lambda(\tau_f)=\infty$, see Fig. \ref{fig:Hawking_instanton}. Such a worldline satisfies the required boundary conditions with $\epsilon=\epsilon'$, and its asymptotic $U$-values of differ by 
\begin{align}
   U(\tau_f)-U(\tau_i)= u-(u-4GM\pi i) =4\pi GM i\;.
\end{align}
Thus we recover the amplitude 
\begin{align}
    \mathcal{A}_{0\rightarrow 2}(\epsilon,\epsilon')\propto \delta(\epsilon-\epsilon')e^{-4\pi GM\epsilon}\,,
\end{align}
which we previously read off from the Feynman modes. The delta function arises because the instanton exists when $\epsilon=\epsilon'$.

\bibliographystyle{JHEP}
\bibliography{WL_schwinger_hawking}

\providecommand{\href}[2]{#2}\begingroup\raggedright\begin{thebibliography}{100}

\bibitem{Travaglini:2022uwo}
G.~Travaglini et~al., \emph{{The SAGEX review on scattering amplitudes}}, \href{http://dx.doi.org/10.1088/1751-8121/ac8380}{\emph{J. Phys. A} {\bfseries 55} (2022) 443001}, [\href{https://arxiv.org/abs/2203.13011}{{\ttfamily 2203.13011}}].

\bibitem{Kosower:2022yvp}
D.~A. Kosower, R.~Monteiro and D.~O'Connell, \emph{{The SAGEX review on scattering amplitudes Chapter 14: Classical gravity from scattering amplitudes}}, \href{http://dx.doi.org/10.1088/1751-8121/ac8846}{\emph{J. Phys. A} {\bfseries 55} (2022) 443015}, [\href{https://arxiv.org/abs/2203.13025}{{\ttfamily 2203.13025}}].

\bibitem{Buonanno:2022pgc}
A.~Buonanno, M.~Khalil, D.~O'Connell, R.~Roiban, M.~P. Solon and M.~Zeng, \emph{{Snowmass White Paper: Gravitational Waves and Scattering Amplitudes}},  in \emph{{Snowmass 2021}}, 4, 2022, \href{https://arxiv.org/abs/2204.05194}{{\ttfamily 2204.05194}}.

\bibitem{Kosower:2018adc}
D.~A. Kosower, B.~Maybee and D.~O'Connell, \emph{{Amplitudes, Observables, and Classical Scattering}}, \href{http://dx.doi.org/10.1007/JHEP02(2019)137}{\emph{JHEP} {\bfseries 02} (2019) 137}, [\href{https://arxiv.org/abs/1811.10950}{{\ttfamily 1811.10950}}].

\bibitem{Cristofoli:2021vyo}
A.~Cristofoli, R.~Gonzo, D.~A. Kosower and D.~O'Connell, \emph{{Waveforms from amplitudes}}, \href{http://dx.doi.org/10.1103/PhysRevD.106.056007}{\emph{Phys. Rev. D} {\bfseries 106} (2022) 056007}, [\href{https://arxiv.org/abs/2107.10193}{{\ttfamily 2107.10193}}].

\bibitem{Kalin:2019rwq}
G.~K\"alin and R.~A. Porto, \emph{{From Boundary Data to Bound States}}, \href{http://dx.doi.org/10.1007/JHEP01(2020)072}{\emph{JHEP} {\bfseries 01} (2020) 072}, [\href{https://arxiv.org/abs/1910.03008}{{\ttfamily 1910.03008}}].

\bibitem{Kalin:2019inp}
G.~K\"alin and R.~A. Porto, \emph{{From boundary data to bound states. Part II. Scattering angle to dynamical invariants (with twist)}}, \href{http://dx.doi.org/10.1007/JHEP02(2020)120}{\emph{JHEP} {\bfseries 02} (2020) 120}, [\href{https://arxiv.org/abs/1911.09130}{{\ttfamily 1911.09130}}].

\bibitem{Cho:2021arx}
G.~Cho, G.~K\"alin and R.~A. Porto, \emph{{From boundary data to bound states. Part III. Radiative effects}}, \href{http://dx.doi.org/10.1007/JHEP04(2022)154}{\emph{JHEP} {\bfseries 04} (2022) 154}, [\href{https://arxiv.org/abs/2112.03976}{{\ttfamily 2112.03976}}].

\bibitem{Adamo:2022ooq}
T.~Adamo and R.~Gonzo, \emph{{Bethe-Salpeter equation for classical gravitational bound states}}, \href{http://dx.doi.org/10.1007/JHEP05(2023)088}{\emph{JHEP} {\bfseries 05} (2023) 088}, [\href{https://arxiv.org/abs/2212.13269}{{\ttfamily 2212.13269}}].

\bibitem{Gonzo:2023goe}
R.~Gonzo and C.~Shi, \emph{{Boundary to bound dictionary for generic Kerr orbits}}, \href{http://dx.doi.org/10.1103/PhysRevD.108.084065}{\emph{Phys. Rev. D} {\bfseries 108} (2023) 084065}, [\href{https://arxiv.org/abs/2304.06066}{{\ttfamily 2304.06066}}].

\bibitem{Adamo:2024oxy}
T.~Adamo, R.~Gonzo and A.~Ilderton, \emph{{Gravitational bound waveforms from amplitudes}}, \href{http://dx.doi.org/10.1007/JHEP05(2024)034}{\emph{JHEP} {\bfseries 05} (2024) 034}, [\href{https://arxiv.org/abs/2402.00124}{{\ttfamily 2402.00124}}].

\bibitem{Buonanno:2024byg}
A.~Buonanno, G.~Mogull, R.~Patil and L.~Pompili, \emph{{Post-Minkowskian Theory Meets the Spinning Effective-One-Body Approach for Bound-Orbit Waveforms}}, \href{http://dx.doi.org/10.1103/PhysRevLett.133.211402}{\emph{Phys. Rev. Lett.} {\bfseries 133} (2024) 211402}, [\href{https://arxiv.org/abs/2405.19181}{{\ttfamily 2405.19181}}].

\bibitem{Aoude:2020ygw}
R.~Aoude, K.~Haddad and A.~Helset, \emph{{Tidal effects for spinning particles}}, \href{http://dx.doi.org/10.1007/JHEP03(2021)097}{\emph{JHEP} {\bfseries 03} (2021) 097}, [\href{https://arxiv.org/abs/2012.05256}{{\ttfamily 2012.05256}}].

\bibitem{Ivanov:2022qqt}
M.~M. Ivanov and Z.~Zhou, \emph{{Vanishing of Black Hole Tidal Love Numbers from Scattering Amplitudes}}, \href{http://dx.doi.org/10.1103/PhysRevLett.130.091403}{\emph{Phys. Rev. Lett.} {\bfseries 130} (2023) 091403}, [\href{https://arxiv.org/abs/2209.14324}{{\ttfamily 2209.14324}}].

\bibitem{Jakobsen:2023pvx}
G.~U. Jakobsen, G.~Mogull, J.~Plefka and B.~Sauer, \emph{{Tidal effects and renormalization at fourth post-Minkowskian order}}, \href{http://dx.doi.org/10.1103/PhysRevD.109.L041504}{\emph{Phys. Rev. D} {\bfseries 109} (2024) L041504}, [\href{https://arxiv.org/abs/2312.00719}{{\ttfamily 2312.00719}}].

\bibitem{Saketh:2023bul}
M.~V.~S. Saketh, Z.~Zhou and M.~M. Ivanov, \emph{{Dynamical tidal response of Kerr black holes from scattering amplitudes}}, \href{http://dx.doi.org/10.1103/PhysRevD.109.064058}{\emph{Phys. Rev. D} {\bfseries 109} (2024) 064058}, [\href{https://arxiv.org/abs/2307.10391}{{\ttfamily 2307.10391}}].

\bibitem{Georgoudis:2023eke}
A.~Georgoudis, C.~Heissenberg and R.~Russo, \emph{{An eikonal-inspired approach to the gravitational scattering waveform}}, \href{http://dx.doi.org/10.1007/JHEP03(2024)089}{\emph{JHEP} {\bfseries 03} (2024) 089}, [\href{https://arxiv.org/abs/2312.07452}{{\ttfamily 2312.07452}}].

\bibitem{Bini:2024rsy}
D.~Bini, T.~Damour, S.~De~Angelis, A.~Geralico, A.~Herderschee, R.~Roiban et~al., \emph{{Gravitational waveforms: A tale of two formalisms}}, \href{http://dx.doi.org/10.1103/PhysRevD.109.125008}{\emph{Phys. Rev. D} {\bfseries 109} (2024) 125008}, [\href{https://arxiv.org/abs/2402.06604}{{\ttfamily 2402.06604}}].

\bibitem{Elkhidir:2024izo}
A.~Elkhidir, D.~O'Connell and R.~Roiban, \emph{{Supertranslations from Scattering Amplitudes}},  \href{https://arxiv.org/abs/2408.15961}{{\ttfamily 2408.15961}}.

\bibitem{Goldberger:2020geb}
W.~D. Goldberger and I.~Z. Rothstein, \emph{{Virtual Hawking Radiation}}, \href{http://dx.doi.org/10.1103/PhysRevLett.125.211301}{\emph{Phys. Rev. Lett.} {\bfseries 125} (2020) 211301}, [\href{https://arxiv.org/abs/2007.00726}{{\ttfamily 2007.00726}}].

\bibitem{Kim:2020dif}
J.-W. Kim and M.~Shim, \emph{{Quantum corrections to tidal Love number for Schwarzschild black holes}}, \href{http://dx.doi.org/10.1103/PhysRevD.104.046022}{\emph{Phys. Rev. D} {\bfseries 104} (2021) 046022}, [\href{https://arxiv.org/abs/2011.03337}{{\ttfamily 2011.03337}}].

\bibitem{Gaddam:2021zka}
N.~Gaddam and N.~Groenenboom, \emph{{2 \textrightarrow{} 2N scattering: Eikonalisation and the Page curve}}, \href{http://dx.doi.org/10.1007/JHEP01(2022)146}{\emph{JHEP} {\bfseries 01} (2022) 146}, [\href{https://arxiv.org/abs/2110.14673}{{\ttfamily 2110.14673}}].

\bibitem{Gaddam:2020mwe}
N.~Gaddam and N.~Groenenboom, \emph{{Soft graviton exchange and the information paradox}}, \href{http://dx.doi.org/10.1103/PhysRevD.109.026007}{\emph{Phys. Rev. D} {\bfseries 109} (2024) 026007}, [\href{https://arxiv.org/abs/2012.02355}{{\ttfamily 2012.02355}}].

\bibitem{Ferreira:2020whz}
R.~Z. Ferreira and C.~Heissenberg, \emph{{Super-Hawking Radiation}}, \href{http://dx.doi.org/10.1007/JHEP02(2021)038}{\emph{JHEP} {\bfseries 02} (2021) 038}, [\href{https://arxiv.org/abs/2011.04688}{{\ttfamily 2011.04688}}].

\bibitem{Melville:2023kgd}
S.~Melville and G.~L. Pimentel, \emph{{de Sitter S matrix for the masses}}, \href{http://dx.doi.org/10.1103/PhysRevD.110.103530}{\emph{Phys. Rev. D} {\bfseries 110} (2024) 103530}, [\href{https://arxiv.org/abs/2309.07092}{{\ttfamily 2309.07092}}].

\bibitem{Aoude:2024sve}
R.~Aoude, D.~O'Connell and M.~Sergola, \emph{{Amplitudes for Hawking Radiation}},  \href{https://arxiv.org/abs/2412.05267}{{\ttfamily 2412.05267}}.

\bibitem{Hartle:1976tp}
J.~B. Hartle and S.~W. Hawking, \emph{{Path Integral Derivation of Black Hole Radiance}}, \href{http://dx.doi.org/10.1103/PhysRevD.13.2188}{\emph{Phys. Rev. D} {\bfseries 13} (1976) 2188--2203}.

\bibitem{Damour:1976jd}
T.~Damour and R.~Ruffini, \emph{{Black Hole Evaporation in the Klein-Sauter-Heisenberg-Euler Formalism}}, \href{http://dx.doi.org/10.1103/PhysRevD.14.332}{\emph{Phys. Rev. D} {\bfseries 14} (1976) 332--334}.

\bibitem{Srinivasan:1998ty}
K.~Srinivasan and T.~Padmanabhan, \emph{{Particle production and complex path analysis}}, \href{http://dx.doi.org/10.1103/PhysRevD.60.024007}{\emph{Phys. Rev. D} {\bfseries 60} (1999) 024007}, [\href{https://arxiv.org/abs/gr-qc/9812028}{{\ttfamily gr-qc/9812028}}].

\bibitem{Shankaranarayanan:2000gb}
S.~Shankaranarayanan, K.~Srinivasan and T.~Padmanabhan, \emph{{Method of complex paths and general covariance of Hawking radiation}}, \href{http://dx.doi.org/10.1142/S0217732301003632}{\emph{Mod. Phys. Lett. A} {\bfseries 16} (2001) 571--578}, [\href{https://arxiv.org/abs/gr-qc/0007022}{{\ttfamily gr-qc/0007022}}].

\bibitem{Shankaranarayanan:2000qv}
S.~Shankaranarayanan, T.~Padmanabhan and K.~Srinivasan, \emph{{Hawking radiation in different coordinate settings: Complex paths approach}}, \href{http://dx.doi.org/10.1088/0264-9381/19/10/310}{\emph{Class. Quant. Grav.} {\bfseries 19} (2002) 2671--2688}, [\href{https://arxiv.org/abs/gr-qc/0010042}{{\ttfamily gr-qc/0010042}}].

\bibitem{Padmanabhan:2004tz}
T.~Padmanabhan, \emph{{Entropy of horizons, complex paths and quantum tunneling}}, \href{http://dx.doi.org/10.1142/S0217732304015257}{\emph{Mod. Phys. Lett. A} {\bfseries 19} (2004) 2637--2643}, [\href{https://arxiv.org/abs/gr-qc/0405072}{{\ttfamily gr-qc/0405072}}].

\bibitem{Shankaranarayanan:2003ya}
S.~Shankaranarayanan, \emph{{Temperature and entropy of Schwarzschild-de Sitter space-time}}, \href{http://dx.doi.org/10.1103/PhysRevD.67.084026}{\emph{Phys. Rev. D} {\bfseries 67} (2003) 084026}, [\href{https://arxiv.org/abs/gr-qc/0301090}{{\ttfamily gr-qc/0301090}}].

\bibitem{Angheben:2005rm}
M.~Angheben, M.~Nadalini, L.~Vanzo and S.~Zerbini, \emph{{Hawking radiation as tunneling for extremal and rotating black holes}}, \href{http://dx.doi.org/10.1088/1126-6708/2005/05/014}{\emph{JHEP} {\bfseries 05} (2005) 014}, [\href{https://arxiv.org/abs/hep-th/0503081}{{\ttfamily hep-th/0503081}}].

\bibitem{Kerner:2006vu}
R.~Kerner and R.~B. Mann, \emph{{Tunnelling, temperature and Taub-NUT black holes}}, \href{http://dx.doi.org/10.1103/PhysRevD.73.104010}{\emph{Phys. Rev. D} {\bfseries 73} (2006) 104010}, [\href{https://arxiv.org/abs/gr-qc/0603019}{{\ttfamily gr-qc/0603019}}].

\bibitem{Parikh:1999mf}
M.~K. Parikh and F.~Wilczek, \emph{{Hawking radiation as tunneling}}, \href{http://dx.doi.org/10.1103/PhysRevLett.85.5042}{\emph{Phys. Rev. Lett.} {\bfseries 85} (2000) 5042--5045}, [\href{https://arxiv.org/abs/hep-th/9907001}{{\ttfamily hep-th/9907001}}].

\bibitem{Hemming:2000as}
S.~Hemming and E.~Keski-Vakkuri, \emph{{Hawking radiation from AdS black holes}}, \href{http://dx.doi.org/10.1103/PhysRevD.64.044006}{\emph{Phys. Rev. D} {\bfseries 64} (2001) 044006}, [\href{https://arxiv.org/abs/gr-qc/0005115}{{\ttfamily gr-qc/0005115}}].

\bibitem{Volovik:2008ww}
G.~E. Volovik, \emph{{On de Sitter radiation via quantum tunneling}}, \href{http://dx.doi.org/10.1142/S0218271809015035}{\emph{Int. J. Mod. Phys. D} {\bfseries 18} (2009) 1227}, [\href{https://arxiv.org/abs/0803.3367}{{\ttfamily 0803.3367}}].

\bibitem{Parikh:2002qh}
M.~K. Parikh, \emph{{New coordinates for de Sitter space and de Sitter radiation}}, \href{http://dx.doi.org/10.1016/S0370-2693(02)02701-6}{\emph{Phys. Lett. B} {\bfseries 546} (2002) 189--195}, [\href{https://arxiv.org/abs/hep-th/0204107}{{\ttfamily hep-th/0204107}}].

\bibitem{Medved:2002zj}
A.~J.~M. Medved, \emph{{Radiation via tunneling from a de Sitter cosmological horizon}}, \href{http://dx.doi.org/10.1103/PhysRevD.66.124009}{\emph{Phys. Rev. D} {\bfseries 66} (2002) 124009}, [\href{https://arxiv.org/abs/hep-th/0207247}{{\ttfamily hep-th/0207247}}].

\bibitem{Wu:2006nj}
S.-Q. Wu and Q.-Q. Jiang, \emph{{Hawking radiation of charged particles as tunneling from higher dimensional Reissner-Nordstrom-de Sitter black holes}},  \href{https://arxiv.org/abs/hep-th/0603082}{{\ttfamily hep-th/0603082}}.

\bibitem{Sauter:1931zz}
F.~Sauter, \emph{{Uber das Verhalten eines Elektrons im homogenen elektrischen Feld nach der relativistischen Theorie Diracs}}, \href{http://dx.doi.org/10.1007/BF01339461}{\emph{Z. Phys.} {\bfseries 69} (1931) 742--764}.

\bibitem{Schwinger:1951nm}
J.~S. Schwinger, \emph{{On gauge invariance and vacuum polarization}}, \href{http://dx.doi.org/10.1103/PhysRev.82.664}{\emph{Phys. Rev.} {\bfseries 82} (1951) 664--679}.

\bibitem{Brezin:1970xf}
E.~Brezin and C.~Itzykson, \emph{{Pair production in vacuum by an alternating field}}, \href{http://dx.doi.org/10.1103/PhysRevD.2.1191}{\emph{Phys. Rev. D} {\bfseries 2} (1970) 1191--1199}.

\bibitem{Fedotov:2022ely}
A.~Fedotov, A.~Ilderton, F.~Karbstein, B.~King, D.~Seipt, H.~Taya et~al., \emph{{Advances in QED with intense background fields}}, \href{http://dx.doi.org/10.1016/j.physrep.2023.01.003}{\emph{Phys. Rept.} {\bfseries 1010} (2023) 1--138}, [\href{https://arxiv.org/abs/2203.00019}{{\ttfamily 2203.00019}}].

\bibitem{Srinivasan:1999ux}
K.~Srinivasan and T.~Padmanabhan, \emph{{A Novel approach to particle production in an uniform electric field}},  \href{https://arxiv.org/abs/gr-qc/9911022}{{\ttfamily gr-qc/9911022}}.

\bibitem{Ilderton:2023ifn}
A.~Ilderton and W.~Lindved, \emph{{Scattering amplitudes and electromagnetic horizons}}, \href{http://dx.doi.org/10.1007/JHEP12(2023)118}{\emph{JHEP} {\bfseries 12} (2023) 118}, [\href{https://arxiv.org/abs/2306.15475}{{\ttfamily 2306.15475}}].

\bibitem{AFFLECK1982509}
I.~K. Affleck, O.~Alvarez and N.~S. Manton, \emph{Pair production at strong coupling in weak external fields}, \href{http://dx.doi.org/https://doi.org/10.1016/0550-3213(82)90455-2}{\emph{Nuclear Physics B} {\bfseries 197} (1982) 509--519}.

\bibitem{Bern:1990cu}
Z.~Bern and D.~A. Kosower, \emph{{Efficient calculation of one loop QCD amplitudes}}, \href{http://dx.doi.org/10.1103/PhysRevLett.66.1669}{\emph{Phys. Rev. Lett.} {\bfseries 66} (1991) 1669--1672}.

\bibitem{Bern:1991aq}
Z.~Bern and D.~A. Kosower, \emph{{The Computation of loop amplitudes in gauge theories}}, \href{http://dx.doi.org/10.1016/0550-3213(92)90134-W}{\emph{Nucl. Phys. B} {\bfseries 379} (1992) 451--561}.

\bibitem{Strassler:1992zr}
M.~J. Strassler, \emph{{Field theory without Feynman diagrams: One loop effective actions}}, \href{http://dx.doi.org/10.1016/0550-3213(92)90098-V}{\emph{Nucl. Phys. B} {\bfseries 385} (1992) 145--184}, [\href{https://arxiv.org/abs/hep-ph/9205205}{{\ttfamily hep-ph/9205205}}].

\bibitem{Edwards:2019eby}
J.~P. Edwards and C.~Schubert, \emph{{Quantum mechanical path integrals in the first quantised approach to quantum field theory}},  12, 2019, \href{https://arxiv.org/abs/1912.10004}{{\ttfamily 1912.10004}}.

\bibitem{Daikouji:1995dz}
K.~Daikouji, M.~Shino and Y.~Sumino, \emph{{Bern-Kosower rule for scalar QED}}, \href{http://dx.doi.org/10.1103/PhysRevD.53.4598}{\emph{Phys. Rev. D} {\bfseries 53} (1996) 4598--4615}, [\href{https://arxiv.org/abs/hep-ph/9508377}{{\ttfamily hep-ph/9508377}}].

\bibitem{Martin:2003gb}
L.~C. Martin, C.~Schubert and V.~M. Villanueva~Sandoval, \emph{{On the low-energy limit of the QED N photon amplitudes}}, \href{http://dx.doi.org/10.1016/S0550-3213(03)00578-9}{\emph{Nucl. Phys. B} {\bfseries 668} (2003) 335--344}, [\href{https://arxiv.org/abs/hep-th/0301022}{{\ttfamily hep-th/0301022}}].

\bibitem{Schubert:2001he}
C.~Schubert, \emph{{Perturbative quantum field theory in the string inspired formalism}}, \href{http://dx.doi.org/10.1016/S0370-1573(01)00013-8}{\emph{Phys. Rept.} {\bfseries 355} (2001) 73--234}, [\href{https://arxiv.org/abs/hep-th/0101036}{{\ttfamily hep-th/0101036}}].

\bibitem{Dunne:2002qf}
G.~V. Dunne and C.~Schubert, \emph{{Two loop selfdual Euler-Heisenberg Lagrangians. 1. Real part and helicity amplitudes}}, \href{http://dx.doi.org/10.1088/1126-6708/2002/08/053}{\emph{JHEP} {\bfseries 08} (2002) 053}, [\href{https://arxiv.org/abs/hep-th/0205004}{{\ttfamily hep-th/0205004}}].

\bibitem{Ahmadiniaz:2020wlm}
N.~Ahmadiniaz, V.~M. Banda~Guzm\'an, F.~Bastianelli, O.~Corradini, J.~P. Edwards and C.~Schubert, \emph{{Worldline master formulas for the dressed electron propagator. Part I. Off-shell amplitudes}}, \href{http://dx.doi.org/10.1007/JHEP08(2020)018}{\emph{JHEP} {\bfseries 08} (2020) 049}, [\href{https://arxiv.org/abs/2004.01391}{{\ttfamily 2004.01391}}].

\bibitem{Ahmadiniaz:2021gsd}
N.~Ahmadiniaz, V.~M.~B. Guzman, F.~Bastianelli, O.~Corradini, J.~P. Edwards and C.~Schubert, \emph{{Worldline master formulas for the dressed electron propagator. Part 2. On-shell amplitudes}}, \href{http://dx.doi.org/10.1007/JHEP01(2022)050}{\emph{JHEP} {\bfseries 01} (2022) 050}, [\href{https://arxiv.org/abs/2107.00199}{{\ttfamily 2107.00199}}].

\bibitem{Gies:2003cv}
H.~Gies, K.~Langfeld and L.~Moyaerts, \emph{{Casimir effect on the worldline}}, \href{http://dx.doi.org/10.1088/1126-6708/2003/06/018}{\emph{JHEP} {\bfseries 06} (2003) 018}, [\href{https://arxiv.org/abs/hep-th/0303264}{{\ttfamily hep-th/0303264}}].

\bibitem{Gies:2006bt}
H.~Gies and K.~Klingmuller, \emph{{Casimir effect for curved geometries: PFA validity limits}}, \href{http://dx.doi.org/10.1103/PhysRevLett.96.220401}{\emph{Phys. Rev. Lett.} {\bfseries 96} (2006) 220401}, [\href{https://arxiv.org/abs/quant-ph/0601094}{{\ttfamily quant-ph/0601094}}].

\bibitem{Shaisultanov:1995tm}
R.~Shaisultanov, \emph{{On the string inspired approach to QED in external field}}, \href{http://dx.doi.org/10.1016/0370-2693(96)00359-0}{\emph{Phys. Lett. B} {\bfseries 378} (1996) 354--356}, [\href{https://arxiv.org/abs/hep-th/9512142}{{\ttfamily hep-th/9512142}}].

\bibitem{Reuter:1996zm}
M.~Reuter, M.~G. Schmidt and C.~Schubert, \emph{{Constant external fields in gauge theory and the spin 0, 1/2, 1 path integrals}}, \href{http://dx.doi.org/10.1006/aphy.1997.5716}{\emph{Annals Phys.} {\bfseries 259} (1997) 313--365}, [\href{https://arxiv.org/abs/hep-th/9610191}{{\ttfamily hep-th/9610191}}].

\bibitem{Schubert:2000yt}
C.~Schubert, \emph{{Vacuum polarization tensors in constant electromagnetic fields. Part 1.}}, \href{http://dx.doi.org/10.1016/S0550-3213(00)00423-5}{\emph{Nucl. Phys. B} {\bfseries 585} (2000) 407--428}, [\href{https://arxiv.org/abs/hep-ph/0001288}{{\ttfamily hep-ph/0001288}}].

\bibitem{Bastianelli:2002fv}
F.~Bastianelli and A.~Zirotti, \emph{{Worldline formalism in a gravitational background}}, \href{http://dx.doi.org/10.1016/S0550-3213(02)00683-1}{\emph{Nucl. Phys. B} {\bfseries 642} (2002) 372--388}, [\href{https://arxiv.org/abs/hep-th/0205182}{{\ttfamily hep-th/0205182}}].

\bibitem{Edwards:2021vhg}
J.~P. Edwards and C.~Schubert, \emph{{N-photon amplitudes in a plane-wave background}}, \href{http://dx.doi.org/10.1016/j.physletb.2021.136696}{\emph{Phys. Lett. B} {\bfseries 822} (2021) 136696}, [\href{https://arxiv.org/abs/2105.08173}{{\ttfamily 2105.08173}}].

\bibitem{Schubert:2023gsl}
C.~Schubert and R.~Shaisultanov, \emph{{Master formulas for photon amplitudes in a combined constant and plane-wave background field}}, \href{http://dx.doi.org/10.1016/j.physletb.2023.137969}{\emph{Phys. Lett. B} {\bfseries 843} (2023) 137969}, [\href{https://arxiv.org/abs/2303.08907}{{\ttfamily 2303.08907}}].

\bibitem{Copinger:2023ctz}
P.~Copinger, J.~P. Edwards, A.~Ilderton and K.~Rajeev, \emph{{Master formulas for N-photon tree level amplitudes in plane wave backgrounds}}, \href{http://dx.doi.org/10.1103/PhysRevD.109.065003}{\emph{Phys. Rev. D} {\bfseries 109} (2024) 065003}, [\href{https://arxiv.org/abs/2311.14638}{{\ttfamily 2311.14638}}].

\bibitem{Copinger:2024twl}
P.~Copinger, J.~P. Edwards, A.~Ilderton and K.~Rajeev, \emph{{All-multiplicity amplitudes in impulsive PP-waves from the worldline formalism}}, \href{http://dx.doi.org/10.1007/JHEP09(2024)148}{\emph{JHEP} {\bfseries 09} (2024) 148}, [\href{https://arxiv.org/abs/2405.07385}{{\ttfamily 2405.07385}}].

\bibitem{Bonezzi:2018box}
R.~Bonezzi, A.~Meyer and I.~Sachs, \emph{{Einstein gravity from the $ \mathcal{N}=4 $ spinning particle}}, \href{http://dx.doi.org/10.1007/JHEP10(2018)025}{\emph{JHEP} {\bfseries 10} (2018) 025}, [\href{https://arxiv.org/abs/1807.07989}{{\ttfamily 1807.07989}}].

\bibitem{Bastianelli:2019xhi}
F.~Bastianelli, R.~Bonezzi, O.~Corradini and E.~Latini, \emph{{One-loop quantum gravity from the $\mathcal N=4$ spinning particle}}, \href{http://dx.doi.org/10.1007/JHEP11(2019)124}{\emph{JHEP} {\bfseries 11} (2019) 124}, [\href{https://arxiv.org/abs/1909.05750}{{\ttfamily 1909.05750}}].

\bibitem{Bastianelli:2013tsa}
F.~Bastianelli and R.~Bonezzi, \emph{{One-loop quantum gravity from a worldline viewpoint}}, \href{http://dx.doi.org/10.1007/JHEP07(2013)016}{\emph{JHEP} {\bfseries 07} (2013) 016}, [\href{https://arxiv.org/abs/1304.7135}{{\ttfamily 1304.7135}}].

\bibitem{Bastianelli:2021nbs}
F.~Bastianelli, F.~Comberiati and L.~de~la Cruz, \emph{{Light bending from eikonal in worldline quantum field theory}}, \href{http://dx.doi.org/10.1007/JHEP02(2022)209}{\emph{JHEP} {\bfseries 02} (2022) 209}, [\href{https://arxiv.org/abs/2112.05013}{{\ttfamily 2112.05013}}].

\bibitem{Mogull:2020sak}
G.~Mogull, J.~Plefka and J.~Steinhoff, \emph{{Classical black hole scattering from a worldline quantum field theory}}, \href{http://dx.doi.org/10.1007/JHEP02(2021)048}{\emph{JHEP} {\bfseries 02} (2021) 048}, [\href{https://arxiv.org/abs/2010.02865}{{\ttfamily 2010.02865}}].

\bibitem{Jakobsen:2021smu}
G.~U. Jakobsen, G.~Mogull, J.~Plefka and J.~Steinhoff, \emph{{Classical Gravitational Bremsstrahlung from a Worldline Quantum Field Theory}}, \href{http://dx.doi.org/10.1103/PhysRevLett.126.201103}{\emph{Phys. Rev. Lett.} {\bfseries 126} (2021) 201103}, [\href{https://arxiv.org/abs/2101.12688}{{\ttfamily 2101.12688}}].

\bibitem{Jakobsen:2021lvp}
G.~U. Jakobsen, G.~Mogull, J.~Plefka and J.~Steinhoff, \emph{{Gravitational Bremsstrahlung and Hidden Supersymmetry of Spinning Bodies}}, \href{http://dx.doi.org/10.1103/PhysRevLett.128.011101}{\emph{Phys. Rev. Lett.} {\bfseries 128} (2022) 011101}, [\href{https://arxiv.org/abs/2106.10256}{{\ttfamily 2106.10256}}].

\bibitem{Jakobsen:2021zvh}
G.~U. Jakobsen, G.~Mogull, J.~Plefka and J.~Steinhoff, \emph{{SUSY in the sky with gravitons}}, \href{http://dx.doi.org/10.1007/JHEP01(2022)027}{\emph{JHEP} {\bfseries 01} (2022) 027}, [\href{https://arxiv.org/abs/2109.04465}{{\ttfamily 2109.04465}}].

\bibitem{Jakobsen:2022fcj}
G.~U. Jakobsen and G.~Mogull, \emph{{Conservative and Radiative Dynamics of Spinning Bodies at Third Post-Minkowskian Order Using Worldline Quantum Field Theory}}, \href{http://dx.doi.org/10.1103/PhysRevLett.128.141102}{\emph{Phys. Rev. Lett.} {\bfseries 128} (2022) 141102}, [\href{https://arxiv.org/abs/2201.07778}{{\ttfamily 2201.07778}}].

\bibitem{Haddad:2024ebn}
K.~Haddad, G.~U. Jakobsen, G.~Mogull and J.~Plefka, \emph{{Spinning bodies in general relativity from bosonic worldline oscillators}}, \href{http://dx.doi.org/10.1007/JHEP02(2025)019}{\emph{JHEP} {\bfseries 02} (2025) 019}, [\href{https://arxiv.org/abs/2411.08176}{{\ttfamily 2411.08176}}].

\bibitem{Driesse:2024feo}
M.~Driesse, G.~U. Jakobsen, A.~Klemm, G.~Mogull, C.~Nega, J.~Plefka et~al., \emph{{Emergence of Calabi\textendash{}Yau manifolds in high-precision black-hole scattering}}, \href{http://dx.doi.org/10.1038/s41586-025-08984-2}{\emph{Nature} {\bfseries 641} (2025) 603--607}, [\href{https://arxiv.org/abs/2411.11846}{{\ttfamily 2411.11846}}].

\bibitem{Ilderton:2014mla}
A.~Ilderton, \emph{{Localisation in worldline pair production and lightfront zero-modes}}, \href{http://dx.doi.org/10.1007/JHEP09(2014)166}{\emph{JHEP} {\bfseries 09} (2014) 166}, [\href{https://arxiv.org/abs/1406.1513}{{\ttfamily 1406.1513}}].

\bibitem{Feldbrugge:2019sew}
J.~L. Feldbrugge, \emph{{Path Integrals in the Sky: Classical and Quantum Problems with Minimal Assumptions}}, Ph.D. thesis, U. Waterloo (main), 2019.

\bibitem{Rajeev:2021zae}
K.~Rajeev, \emph{{Lorentzian worldline path integral approach to Schwinger effect}}, \href{http://dx.doi.org/10.1103/PhysRevD.104.105014}{\emph{Phys. Rev. D} {\bfseries 104} (2021) 105014}, [\href{https://arxiv.org/abs/2105.12194}{{\ttfamily 2105.12194}}].

\bibitem{Furry:1951bef}
W.~H. Furry, \emph{{On Bound States and Scattering in Positron Theory}}, \href{http://dx.doi.org/10.1103/PhysRev.81.115}{\emph{Phys. Rev.} {\bfseries 81} (1951) 115}.

\bibitem{DeWitt:1967ub}
B.~S. DeWitt, \emph{{Quantum Theory of Gravity. 2. The Manifestly Covariant Theory}}, \href{http://dx.doi.org/10.1103/PhysRev.162.1195}{\emph{Phys. Rev.} {\bfseries 162} (1967) 1195--1239}.

\bibitem{tHooft:1975uxh}
G.~'t~Hooft, \emph{{The Background Field Method in Gauge Field Theories}},  in \emph{{12th Annual Winter School of Theoretical Physics}}, 1975.

\bibitem{Abbott:1981ke}
L.~F. Abbott, \emph{{Introduction to the Background Field Method}}, {\emph{Acta Phys. Polon. B} {\bfseries 13} (1982) 33}.

\bibitem{Hawking:1975vcx}
S.~W. Hawking, \emph{{Particle Creation by Black Holes}}, \href{http://dx.doi.org/10.1007/BF02345020}{\emph{Commun. Math. Phys.} {\bfseries 43} (1975) 199--220}.

\bibitem{Tomaras:2000ag}
T.~N. Tomaras, N.~C. Tsamis and R.~P. Woodard, \emph{{Back reaction in light cone QED}}, \href{http://dx.doi.org/10.1103/PhysRevD.62.125005}{\emph{Phys. Rev. D} {\bfseries 62} (2000) 125005}, [\href{https://arxiv.org/abs/hep-ph/0007166}{{\ttfamily hep-ph/0007166}}].

\bibitem{Tomaras:2001vs}
T.~N. Tomaras, N.~C. Tsamis and R.~P. Woodard, \emph{{Pair creation and axial anomaly in light cone QED(2)}}, \href{http://dx.doi.org/10.1088/1126-6708/2001/11/008}{\emph{JHEP} {\bfseries 11} (2001) 008}, [\href{https://arxiv.org/abs/hep-th/0108090}{{\ttfamily hep-th/0108090}}].

\bibitem{Ahmad:2016vvw}
A.~Ahmad, N.~Ahmadiniaz, O.~Corradini, S.~P. Kim and C.~Schubert, \emph{{Master formulas for the dressed scalar propagator in a constant field}}, \href{http://dx.doi.org/10.1016/j.nuclphysb.2017.03.007}{\emph{Nucl. Phys. B} {\bfseries 919} (2017) 9--24}, [\href{https://arxiv.org/abs/1612.02944}{{\ttfamily 1612.02944}}].

\bibitem{Kim:2000un}
S.~P. Kim and D.~N. Page, \emph{{Schwinger pair production via instantons in a strong electric field}}, \href{http://dx.doi.org/10.1103/PhysRevD.65.105002}{\emph{Phys. Rev. D} {\bfseries 65} (2002) 105002}, [\href{https://arxiv.org/abs/hep-th/0005078}{{\ttfamily hep-th/0005078}}].

\bibitem{Dunne:2005sx}
G.~V. Dunne and C.~Schubert, \emph{{Worldline instantons and pair production in inhomogeneous fields}}, \href{http://dx.doi.org/10.1103/PhysRevD.72.105004}{\emph{Phys. Rev. D} {\bfseries 72} (2005) 105004}, [\href{https://arxiv.org/abs/hep-th/0507174}{{\ttfamily hep-th/0507174}}].

\bibitem{Dunne:2006st}
G.~V. Dunne, Q.-h. Wang, H.~Gies and C.~Schubert, \emph{{Worldline instantons. II. The Fluctuation prefactor}}, \href{http://dx.doi.org/10.1103/PhysRevD.73.065028}{\emph{Phys. Rev. D} {\bfseries 73} (2006) 065028}, [\href{https://arxiv.org/abs/hep-th/0602176}{{\ttfamily hep-th/0602176}}].

\bibitem{Dunne:2006ur}
G.~V. Dunne and Q.-h. Wang, \emph{{Multidimensional Worldline Instantons}}, \href{http://dx.doi.org/10.1103/PhysRevD.74.065015}{\emph{Phys. Rev. D} {\bfseries 74} (2006) 065015}, [\href{https://arxiv.org/abs/hep-th/0608020}{{\ttfamily hep-th/0608020}}].

\bibitem{Dumlu:2011cc}
C.~K. Dumlu and G.~V. Dunne, \emph{{Complex Worldline Instantons and Quantum Interference in Vacuum Pair Production}}, \href{http://dx.doi.org/10.1103/PhysRevD.84.125023}{\emph{Phys. Rev. D} {\bfseries 84} (2011) 125023}, [\href{https://arxiv.org/abs/1110.1657}{{\ttfamily 1110.1657}}].

\bibitem{Ilderton:2015qda}
A.~Ilderton, G.~Torgrimsson and J.~W\r{a}rdh, \emph{{Nonperturbative pair production in interpolating fields}}, \href{http://dx.doi.org/10.1103/PhysRevD.92.065001}{\emph{Phys. Rev. D} {\bfseries 92} (2015) 065001}, [\href{https://arxiv.org/abs/1506.09186}{{\ttfamily 1506.09186}}].

\bibitem{Dumlu:2017kfp}
C.~K. Dumlu, \emph{{Hawking Radiation via Complex Geodesics}}, \href{http://dx.doi.org/10.1103/PhysRevD.98.045019}{\emph{Phys. Rev. D} {\bfseries 98} (2018) 045019}, [\href{https://arxiv.org/abs/1710.07644}{{\ttfamily 1710.07644}}].

\bibitem{DegliEsposti:2021its}
G.~Degli~Esposti and G.~Torgrimsson, \emph{{Worldline instantons for nonlinear Breit-Wheeler pair production and Compton scattering}}, \href{http://dx.doi.org/10.1103/PhysRevD.105.096036}{\emph{Phys. Rev. D} {\bfseries 105} (2022) 096036}, [\href{https://arxiv.org/abs/2112.11433}{{\ttfamily 2112.11433}}].

\bibitem{DegliEsposti:2022yqw}
G.~Degli~Esposti and G.~Torgrimsson, \emph{{Worldline instantons for the momentum spectrum of Schwinger pair production in spacetime dependent fields}}, \href{http://dx.doi.org/10.1103/PhysRevD.107.056019}{\emph{Phys. Rev. D} {\bfseries 107} (2023) 056019}, [\href{https://arxiv.org/abs/2212.11578}{{\ttfamily 2212.11578}}].

\bibitem{DegliEsposti:2023qqu}
G.~Degli~Esposti and G.~Torgrimsson, \emph{{Momentum spectrum of Schwinger pair production in four-dimensional e-dipole fields}}, \href{http://dx.doi.org/10.1103/PhysRevD.109.016013}{\emph{Phys. Rev. D} {\bfseries 109} (2024) 016013}, [\href{https://arxiv.org/abs/2308.01659}{{\ttfamily 2308.01659}}].

\bibitem{DegliEsposti:2024upq}
G.~Degli~Esposti and G.~Torgrimsson, \emph{{Schwinger pair production in spacetime fields: Moir\'e patterns, Aharonov-Bohm phases and Sturm-Liouville eigenvalues}},  \href{https://arxiv.org/abs/2412.19709}{{\ttfamily 2412.19709}}.

\bibitem{Lippstreu:2025jit}
L.~Lippstreu, \emph{{Analytic Properties of Infrared-Finite Amplitudes in Theories with Long-Range Forces}},  \href{https://arxiv.org/abs/2505.04702}{{\ttfamily 2505.04702}}.

\bibitem{Fradkin1991QED}
E.~S. Fradkin, D.~M. Gitman and S.~M. Shvartsman, \emph{Quantum Electrodynamics with Unstable Vacuum}.
\newblock Springer Series in Nuclear and Particle Physics. Springer-Verlag, Berlin, 1991.

\bibitem{Wald1994QFT}
R.~M. Wald, \emph{Quantum Field Theory in Curved Spacetime and Black Hole Thermodynamics}.
\newblock Chicago Lectures in Physics. University of Chicago Press, Chicago, 1994.

\bibitem{Copinger:2024pai}
P.~Copinger, J.~P. Edwards, A.~Ilderton and K.~Rajeev, \emph{{Pair creation, backreaction, and resummation in strong fields}}, \href{http://dx.doi.org/10.1103/PhysRevD.111.036009}{\emph{Phys. Rev. D} {\bfseries 111} (2025) 036009}, [\href{https://arxiv.org/abs/2411.06203}{{\ttfamily 2411.06203}}].

\bibitem{ToAppear}
R.~Aoude, A.~Elkhidir, A.~Ilderton, D.~O'Connell and K.~Rajeev{\emph{{, to appear\!}} }.

\bibitem{Unruh:2004zk}
W.~G. Unruh and R.~Schutzhold, \emph{{On the universality of the Hawking effect}}, \href{http://dx.doi.org/10.1103/PhysRevD.71.024028}{\emph{Phys. Rev. D} {\bfseries 71} (2005) 024028}, [\href{https://arxiv.org/abs/gr-qc/0408009}{{\ttfamily gr-qc/0408009}}].

\bibitem{Parker:1979mf}
L.~Parker, \emph{{Path integrals for a particle in curved space}}, \href{http://dx.doi.org/10.1103/PhysRevD.19.438}{\emph{Phys. Rev. D} {\bfseries 19} (1979) 438--441}.

\bibitem{Bastianelli_vanNieuwenhuizen_2006}
F.~Bastianelli and P.~van Nieuwenhuizen, \emph{Path Integrals and Anomalies in Curved Space}.
\newblock Cambridge Monographs on Mathematical Physics. Cambridge University Press, 2006.

\bibitem{Bastianelli:1991be}
F.~Bastianelli, \emph{{The Path integral for a particle in curved spaces and Weyl anomalies}}, \href{http://dx.doi.org/10.1016/0550-3213(92)90070-R}{\emph{Nucl. Phys. B} {\bfseries 376} (1992) 113--126}, [\href{https://arxiv.org/abs/hep-th/9112035}{{\ttfamily hep-th/9112035}}].

\bibitem{kleinert2006path}
H.~Kleinert, \emph{Path Integrals in Quantum Mechanics, Statistics, Polymer Physics, and Financial Markets}.
\newblock World Scientific Publishing Company, Singapore, 5~ed., 2006.

\bibitem{chaichian2018path}
M.~Chaichian and A.~Demichev, \emph{Path Integrals in Physics: Volume I Stochastic Processes and Quantum Mechanics}.
\newblock CRC Press, Boca Raton, FL, 2018.

\bibitem{Vassilevich:2003xt}
D.~V. Vassilevich, \emph{{Heat kernel expansion: User's manual}}, \href{http://dx.doi.org/10.1016/j.physrep.2003.09.002}{\emph{Phys. Rept.} {\bfseries 388} (2003) 279--360}, [\href{https://arxiv.org/abs/hep-th/0306138}{{\ttfamily hep-th/0306138}}].

\bibitem{Bastianelli:2006hq}
F.~Bastianelli, O.~Corradini and P.~A.~G. Pisani, \emph{{Worldline approach to quantum field theories on flat manifolds with boundaries}}, \href{http://dx.doi.org/10.1088/1126-6708/2007/02/059}{\emph{JHEP} {\bfseries 02} (2007) 059}, [\href{https://arxiv.org/abs/hep-th/0612236}{{\ttfamily hep-th/0612236}}].

\bibitem{Bastianelli:2008vh}
F.~Bastianelli, O.~Corradini, P.~A.~G. Pisani and C.~Schubert, \emph{{Scalar heat kernel with boundary in the worldline formalism}}, \href{http://dx.doi.org/10.1088/1126-6708/2008/10/095}{\emph{JHEP} {\bfseries 10} (2008) 095}, [\href{https://arxiv.org/abs/0809.0652}{{\ttfamily 0809.0652}}].

\bibitem{Asorey:2007zza}
M.~Asorey, J.~Clemente-Gallardo and J.~M. Munoz-Castaneda, \emph{{Boundary conditions: The path integral approach}}, \href{http://dx.doi.org/10.1088/1742-6596/87/1/012004}{\emph{J. Phys. Conf. Ser.} {\bfseries 87} (2007) 012004}, [\href{https://arxiv.org/abs/0712.4353}{{\ttfamily 0712.4353}}].

\bibitem{Corradini:2019nbb}
O.~Corradini, J.~P. Edwards, I.~Huet, L.~Manzo and P.~Pisani, \emph{{Worldline formalism for a confined scalar field}}, \href{http://dx.doi.org/10.1007/JHEP08(2019)037}{\emph{JHEP} {\bfseries 08} (2019) 037}, [\href{https://arxiv.org/abs/1905.00945}{{\ttfamily 1905.00945}}].

\bibitem{Li:2016sjq}
W.-D. Li, Y.-Z. Chen and W.-S. Dai, \emph{{Scattering state and bound state of scalar field in Schwarzschild spacetime: Exact solution}}, \href{http://dx.doi.org/10.1016/j.aop.2019.167919}{\emph{Annals Phys.} {\bfseries 409} (2019) 167919}, [\href{https://arxiv.org/abs/1612.02644}{{\ttfamily 1612.02644}}].

\bibitem{GregerNew}
P.~Semr\'en and G.~Torgrimsson, \emph{{\it Worldline instantons for nonperturbative particle production by space and time dependent gravitational fields}}, {\emph{to appear\!\!} }.

\bibitem{Barut:1989mc}
A.~O. Barut and I.~H. Duru, \emph{{Pair production in an electric field in a time dependent gauge}}, \href{http://dx.doi.org/10.1103/PhysRevD.41.1312}{\emph{Phys. Rev. D} {\bfseries 41} (1990) 1312}.

\end{thebibliography}\endgroup

\end{document}